\documentclass[10pt, a4paper,aps,prl,superscriptaddress,floatfix,twocolumn]{revtex4-2}
 \usepackage{epsfig}
 \usepackage{latexsym}
 \usepackage{epic}
 \usepackage{eepic}
 \usepackage{psfrag}
 \usepackage{rotating}
 \usepackage{multirow,booktabs}
 \usepackage{color}
 \usepackage{cprotect} %verbatim in caption
 \usepackage{miller} %for miller indices
 \usepackage{xspace} %for xspace given a space after ensuremath macro

\usepackage{graphicx}
 
\usepackage{gensymb} %\degree
 \usepackage{amssymb,amsfonts,amsthm,amsmath}
  \usepackage{stmaryrd} % für blitz \fail
  \usepackage{import}

  \usepackage{array}

  \usepackage[normalem]{ulem}

\input{MyBibalias.tex}
\renewcommand\vec{\mathbf}

\newcommand{\dd}{\ensuremath{\rm d}\xspace}
\newcommand{\kk}{\ensuremath{\vec{k}}\xspace}
\newcommand{\nkk}{\ensuremath{|\vec{k}|}\xspace}
\newcommand{\nkkp}{\ensuremath{|\vec{k}'|}\xspace}

\newcommand{\uu}{\ensuremath{\vec{u}}\xspace}
\newcommand{\uuh}{\ensuremath{\widehat{\vec{u}}}\xspace}

\newcommand{\ww}{\ensuremath{\vec{\omega}}\xspace}
\newcommand{\wwh}{\ensuremath{\widehat{\vec{\omega}}}\xspace}

\newcommand{\nueff}{\ensuremath{\nu_{\mathrm{eff}}}\xspace}
\newcommand{\xx}{\ensuremath{\vec{x}}\xspace}

\newcommand{\vl}{\ensuremath{\vec{l}}\xspace}

\newcommand{\GO}{\ensuremath{\Gamma_0}\xspace}
\newcommand{\GT}{\ensuremath{\Gamma_2}\xspace}
\newcommand{\GF}{\ensuremath{\Gamma_4}\xspace}

\newcommand{\Lap}{\ensuremath{\Delta}\xspace}

\usepackage{grffile}

\begin{document}

\title{Nonequilibrium hyperuniform states in active turbulence}

\author{R. Backofen}
\affiliation{Institute of Scientific Computing, TU Dresden, 01062 Dresden, Germany}

\author{Abdelrahman Y. A. Altawil}
\affiliation{Institute of Scientific Computing, TU Dresden, 01062 Dresden, Germany}

\author{Marco Salvalaglio} 
\affiliation{Institute of Scientific Computing, TU Dresden, 01062 Dresden, Germany}
\affiliation{Dresden Centre for Computational Materials Science (DCMS), TU Dresden, 01062 Dresden, Germany}

\author{A. Voigt}
\affiliation{Institute of Scientific Computing, TU Dresden, 01062 Dresden, Germany}
\affiliation{Dresden Centre for Computational Materials Science (DCMS), TU Dresden, 01062 Dresden, Germany} 
\affiliation{Center for Systems Biology Dresden, Pfotenhauerstr. 108, 01307 Dresden, Germany}
\affiliation{Cluster of Excellence, Physics of Life, TU Dresden, Arnoldstr. 18, 01307 Dresden, Germany}

% Keywords are not mandatory, but authors are strongly encouraged to provide them. If provided, please include two to five keywords, separated by the pipe symbol, e.g:

\begin{abstract}
We demonstrate that the complex spatiotemporal structure in active fluids can feature characteristics of hyperuniformity. Using a hydrodynamic model, we show that the transition from hyperuniformity to non-hyperuniformity and anti-hyperuniformity depends on the strength of active forcing and can be related to features of active turbulence without and with scaling characteristics of inertial turbulence. Combined with identified signatures of Levy walks and non-universal diffusion in these systems, this allows for a biological interpretation and the speculation of non-equilibrium hyperuniform states in active fluids as optimal states with respect to robustness and strategies of evasion and foraging.
\end{abstract}

\maketitle

\section{Introduction}
Disordered hyperuniform systems show hybrid crystal-liquid characteristics which endows them with unique isotropic properties and robustness against defects \cite{TS03,Tor18}. Perfect crystals exhibit both short- and long-range correlations, which lead to a structure factor $\widehat{S}(|\mathbf{k}|)$ with sharp, ordered peaks and $\widehat{S}(|\mathbf{k}|) \to 0$ as $|\mathbf{k}| \to 0$, where $\mathbf{k}$ being the wavevector and $|\mathbf{k}|$ the wavenumber. Equivalently, the local number variance $\sigma^2(R)$ in perfect crystals scales with the window size of observation $R$ as $\sigma^2(R) \sim R^{\beta}$ with $\beta = d-1$, where $d$ is the dimensionality of the system. Contrary to this, for conventional liquids, $\widehat{S}(|\mathbf{k}|) \to \text{const.} > 0$ as $|\mathbf{k}| \to 0$ and $\sigma^2(R) \sim R^{\beta}$ with $\beta = d$. Hyperuniform systems are characterized by $\widehat{S}(|\mathbf{k}|) \sim |\mathbf{k}|^{\alpha} $ with $\alpha > 0$ and $d-1 \leq \beta < d$ for $|\mathbf{k}| \to 0$, with different classifications and types depending on $\alpha$ and $\beta$ \cite{Tor18}. Perfect crystals belong to these systems. However, disordered systems with these characteristics also exist and have been identified in physics \cite{Wilkenetal_PRL_2021,SanchezPRB2023}, material science \cite{Zhengetal_SA_2020,Chenetal_PNAS_2021} and mathematics \cite{Ghoshetal_JSP_2017,Chatterjee_PTRF_2019}, to name just a few, and have been shown to be advantageous in technological applications, e.g., photonic crystals with full isotropic bandgaps 
\cite{Fetal2009,froufe2017band,mitchell2018amorphous,gerasimenko2019quantum,yu2021engineered,Vynck2023} or solar cells with increased absorption rates \cite{Tavakoli_ACS_2022}. Less explored is the concept of hyperuniformity in living systems. Examples are the photoreceptor mosaic in the avian retina \cite{Jiaoetal_PRE_2014} or the repertoire of lymphocyte receptors in the adaptive immune system \cite{Mayeretal_PNAS_2015}, where disordered hyperuniformity leads to nearly optimal functionality. Active systems, which comprise individually driven units that convert locally stored energy into mechanical motion \cite{RamaswanyARCMP2010,Marchetti2013}, are conventionally attributed with giant number fluctuations \cite{Ramaswamyetal_EPL_2003,Palacci2013,Alaimoetal_NJP_2016} or motility induced phase separation \cite{Buttinonietal_PRL_2013,Catesetal_ARCMP_2015} for which $\beta > d$ and $\widehat{S}(|\mathbf{k}|) \to \infty$ as $|\mathbf{k}| \to 0$ and thus hyperuniformity is prohibited. Such properties have even been termed "anti-hyperuniformity" \cite{Tor18}. However, characteristics of hyperuniformity have also been identified in active systems. The few examples in this context are \cite{Leietal_PNAS_2019,Leietal_SA_2019,Huangetal_PNAS_2021,Oppenheimeretal_NC_2022}. The required long-range correlation in these systems results from the generated fluid flow, which leads to effective repulsion in the far field and thus suppressed density fluctuations. The experimental system in \cite{Huangetal_PNAS_2021} considers the collective dynamics of marine algae Effrenium voratum, which swim in circles on an air-liquid interface where they self-organize into disordered hyperuniform states. The behaviour has been suggested by two-dimensional particle simulations \cite{Leietal_PNAS_2019,Leietal_SA_2019,Huangetal_PNAS_2021,Oppenheimeretal_NC_2022} and analysed by analytical treatments of modified Navier-Stokes equations \cite{Leietal_PNAS_2019} and an active polar gel model \cite{Leietal_SA_2019}. The discrepancy with conventional models for active systems, which prohibit hyperuniformity, results from neglected hydrodynamic interactions in these models. 

We consider a momentum-conserving extension of the Navier-Stokes equations, known as the generalized Navier-Stokes (GNS) equations \cite{Slomkaetal_EPJSP_2015} and demonstrate using two-dimensional simulations that for low activities, such systems form a non-equilibrium hyperuniform fluid phase. The found transition between hyperuniformity ($\widehat{S}(|\mathbf{k}|) \to 0$) and non-hyperuniformity ($\widehat{S}(|\mathbf{k}|) \to \text{const.} > 0$) or even "anti-hyperuniformity" ($\widehat{S}(|\mathbf{k}|) \to \infty$) \cite{Tor18} as $|\mathbf{k}| \to 0$ thereby relates to the transition between active turbulence states without and with characteristics of inertial turbulence \cite{Linkmannetal_PRL_2019}. The considered simulations address the basic ingredients for non-equilibrium hyperuniformity discussed in \cite{Leietal_PNAS_2019,Leietal_SA_2019,Huangetal_PNAS_2021,Oppenheimeretal_NC_2022}. While being restricted to $d = 2$, they consider hydrodynamic interactions and nonlinear chiral motion. We first review the properties of active turbulence and describe the GNS equations, before computational results are tested against various properties and discussed, and finally conclusions and biological interpretations are given.

\section{Active versus inertial turbulence}
Active systems are prone to experience instabilities and self-organization phenomena, thus developing correlated collective flows that can become spatiotemporally chaotic. This makes them analogous to inertial turbulence at a descriptive level, and hence, these flows are commonly termed active turbulence \cite{Alertetal_ARCMP_2022}. A huge effort has been made to identify similarities and discrepancies between well-explored inertial turbulence and the potentially new active turbulence \cite{Bratanovetal_PNAS_2015,Mukherjeeetal_NP_2023}. Classical inertial fluids undergo a laminar-to-turbulent transition at moderately large Reynolds numbers. Above this transition, inertial effects destabilize the flow and lead to chaotic patterns of vortices and jets. At fully developed inertial turbulence, external driving injects kinetic energy at a scale at which viscous dissipation is negligible. Inertial effects transport this energy across scales until it dissipates by viscous effects at large scales. In the intermediate scales of this energy cascade, the flow exhibits self-similarity, and velocity correlations are scale-invariant. For this regime, the energy spectrum follows a power law $\widehat{E}(|\mathbf{k}|) \sim |\mathbf{k}|^{-5/3}$, with a universal exponent independent of the external driving and the properties of the fluid \cite{Kolmogorov}. %For systems confined to be quasi-two-dimensional a dual energy cascade emerges with leads to different scaling exponents \cite{Boffettaetal_ARFM_2012}. 
Even if such scale separation is not present in active flows, power laws have also been found in active turbulence \cite{Bratanovetal_PNAS_2015,Linkmannetal_PRL_2019,Mukherjeeetal_NP_2023}, if the activity is above some threshold. Simulations also suggest that the flow exhibits spatio-temporal intermittency above this threshold \cite{Mukherjeeetal_NP_2023}, which further supports the connection to inertial turbulence. However, differences have also been established for this regime. In \cite{Bratanovetal_PNAS_2015} it is shown that the observed scaling laws are parameter dependent, which demonstrates the non-universal behaviour of active turbulence. However, this depends on the considered model; within the context of active nematodynamics universal scaling is suggested \cite{Giomi_PRX_2015,Alertetal_NP_2020}. Within a Lagrangian perspective, the mean-squared displacements (MSD) of tracer trajectories in inertial turbulence have a universal behavior of diffusive self-separation \cite{Xiaetal_NC_2013}. Experiments and simulations in active turbulence show signatures of Levy walks and non-universal diffusion \cite{Kurtulduetal_PNAS_2011,Morozovetal_SM_2014,Arieletal_NC_2015,Mukherjeeetal_PRL_2021,Singhetal_PRF_2022}. In \cite{Mukherjeeetal_PRL_2021} this behavior is attributed to oscillatory streaks emerging at large active driving. Given that Lévy walks increase the encounter probability in stochastic
search \cite{Viswanathanetal_Nature_1999,Bartumeusetal_PRL_2002,Humphriesetal_PNAS_2012}, this suggests an optimal search strategy at large activities. Most of these comparisons between active and inertial turbulence focus on intermediate scales \cite{Alertetal_ARCMP_2022} and therefore do not address hyperuniformity, which is a property at large spatial scales.

\section{Modeling active turbulence} \label{sec:1b}
\begin{figure}[htb]
  \noindent
  \begin{tabular}{c}
  \includegraphics*[width = 0.45 \textwidth ]{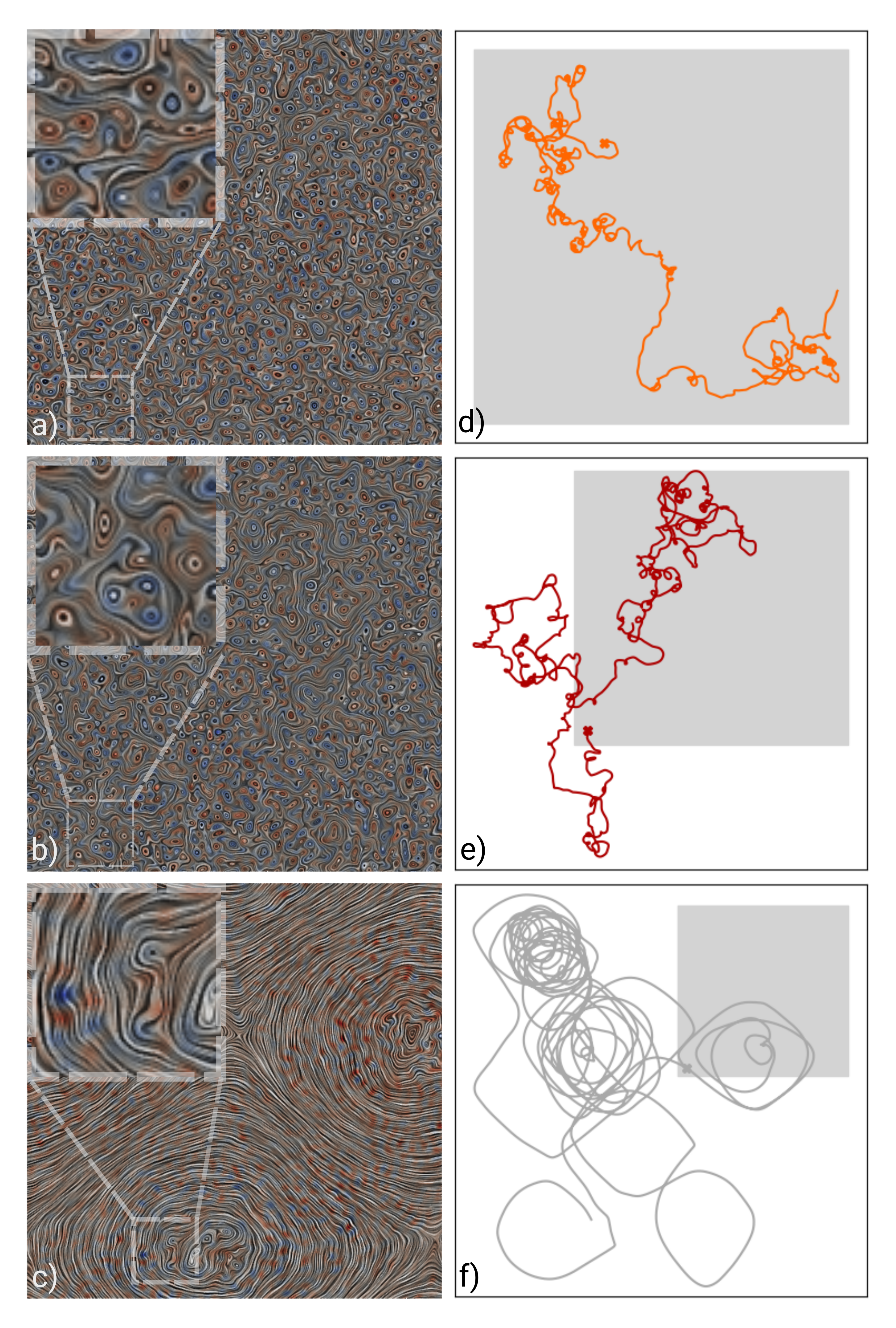}
  \end{tabular}
%  \begin{tabular}{ll}
%    \includegraphics*[width = 0.21 \textwidth ]{figures/LIC_1_crop}
%    &\includegraphics*[width = 0.21 \textwidth ]{figures/sgPartPath0_1_crop}
%    \\
%    a) & d) \\
%    \includegraphics*[width = 0.21 \textwidth ]{figures/LIC_2_crop} 
%    &\includegraphics*[width = 0.21 \textwidth ]{figures/sgPartPath0_2_crop}
 %   \\
 %   b) & e) \\
 %   \includegraphics*[width = 0.21 \textwidth ]{figures/LIC_2p5_crop} 
 %      &\includegraphics*[width = 0.21 \textwidth ]%{{figures/sgPartPath0_2.5_crop}}\\
 %   c) & f)     
  %\end{tabular}
  \begin{center}
\begin{minipage}{0.49\textwidth}
  \caption[short figure description]{
a) - c) Snapshots of flow fields in the periodic domain within the statistically steady states. LIC visualisation with color coding corresponding to vorticity $\ww$. Corresponding movies are provided in SI, a) $- \Gamma_2/\Gamma_0 = 1$, b) $- \Gamma_2/\Gamma_0 = 2$ and c) $- \Gamma_2/\Gamma_0 = 2.5$. The corresponding inlets highlight the microvortices in these flows. d) - f) Corresponding trajectories of a tracer particle in the flow fields shown in the corresponding movies to a) - c). The scaled shaded region corresponds to the considered periodic domain.
\label{fig1}
}
\end{minipage}
\end{center}
\end{figure}

We consider the GNS equations \cite{Slomkaetal_EPJSP_2015}, which relate to the phenomenologically-proposed Toner-Tu-Swift-Hohenberg (TTSH) equations \cite{Wensinketal_PNAS_2012,Dunkeletal_PRL_2013} for bacterial suspensions. However, the TTSH equations, also considered in \cite{Bratanovetal_PNAS_2015,Mukherjeeetal_PRL_2021,Mukherjeeetal_NP_2023} neglect hydrodynamic interactions mediated by the solvent, which makes it a dry system \cite{Alertetal_ARCMP_2022}. In the GNS equations
\begin{align} 
\partial_t \uu + \uu \cdot \nabla \uu + \nabla p = \nabla \cdot \boldsymbol{\sigma}, \quad \nabla \cdot \uu = 0 \label{eq:GNS}
\end{align}
with velocity $\uu(\mathbf{x},t)$, pressure $p(\mathbf{x},t)$ and effective stress tensor $\boldsymbol{\sigma}(\mathbf{x},t)$ which comprises passive contributions from the intrinsic fluid viscosity and active contributions representing the forces exerted by the bacteria on the fluid. It reads
\begin{align}
\sigma_{ij} = \nueff \left( \partial_i u_j + \partial_j u_i \right) \; \mbox{with} \;\;
\nueff = \GO - \GT \Lap + \GF \Lap^2
\end{align}
and $\GO >0$, $\GT < 0$ and $\GF > 0$. These parameters are the simplest choice of an isotropic active stress tensor \cite{Slomkaetal_PRF_2017}, which produces a band of unstable modes injecting energy to drive flows and produce patterns of vortices. This can become turbulent due to the nonlinear advection, as in inertial turbulence. $\GO$ describes the damping of long-wavelength perturbations on scales much larger than the typical correlation length of the coherent flow structures, whereas $\GT$ and $\GF$ account for the growth and damping of modes at intermediate and small scales. For suitably chosen values, \eqref{eq:GNS} reproduces the experimentally observed bulk vortex dynamics of bacterial suspensions \cite{Sokolovetal_PRL_2012}. We explore this model numerically using an approach similar to \cite{Linkmannetal_PRL_2019,Linkmannetal_PRE_2020,Linkmannetal_JFM_2020} and approximate the effective viscosity by a piecewise constant function in reciprocal space, see Section \ref{sec:4}. Figure \ref{fig1} shows snapshots of the flow field within statistically steady states together with corresponding trajectories of tracer particles for selected forcing parameters $- \Gamma_2/\Gamma_0$.

\section{Results}
\label{sec:2}

\begin{figure}[htb]
  \noindent
  \begin{tabular}{l}
    \includegraphics*[width = 0.45 \textwidth ]{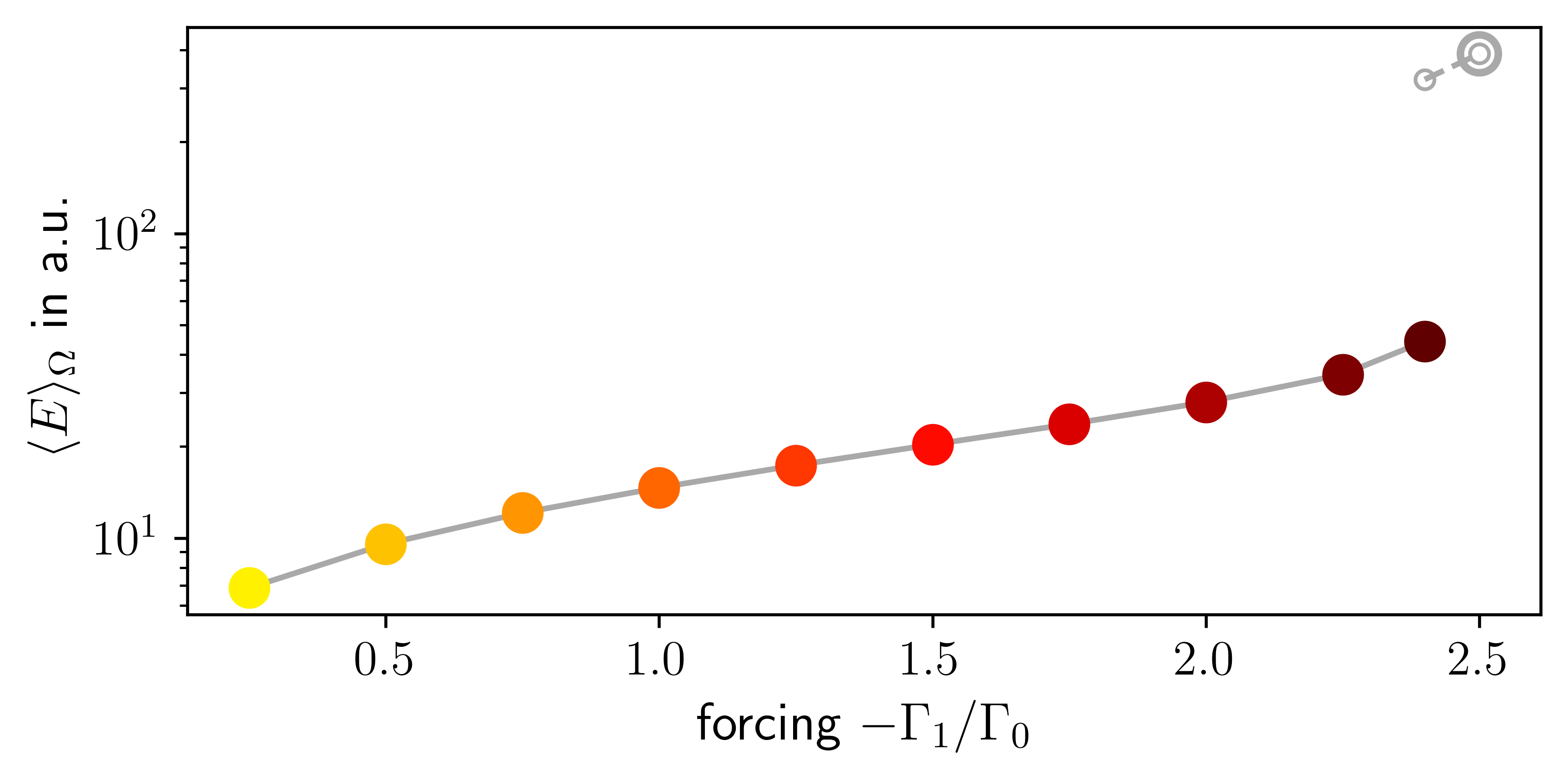}
    \end{tabular}
  \begin{center}
    \begin{minipage}{0.45\textwidth}
      \caption[short figure description]{Time-averaged spatial mean energy of statistically steady-states  $\langle E \rangle_\Omega$ vs. forcing $ |\Gamma_2/\Gamma_0|$. The colors are chosen according to the forcing and are used consistently in all figures. Open symbols indicate condensate states. 
        \label{fig2}
        }
\end{minipage}
\end{center}
\end{figure}

\begin{figure}[htb]
  \noindent
  \begin{tabular}{l}
    \includegraphics*[width = 0.45 \textwidth ]{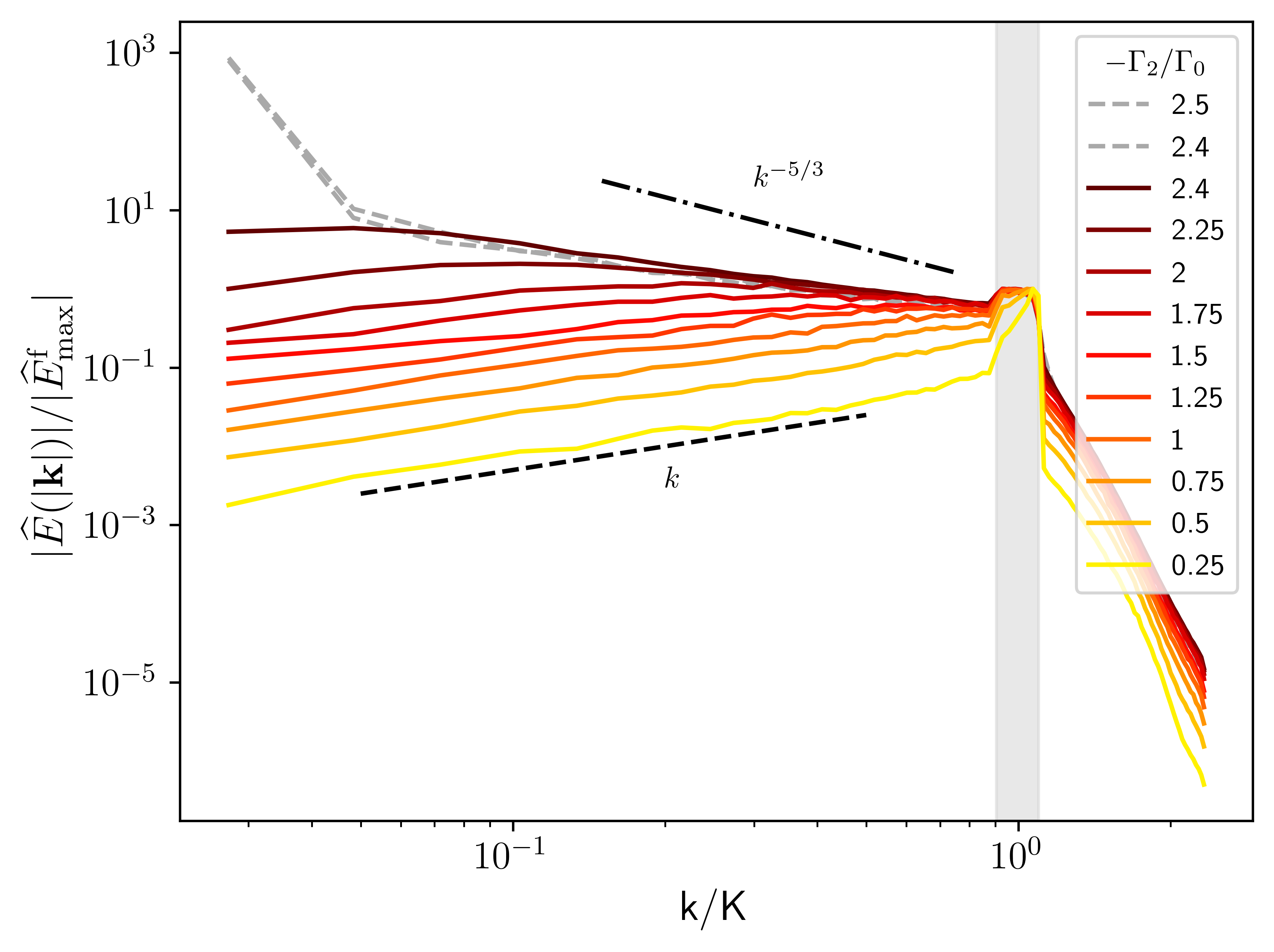}
  \end{tabular}
  \begin{center}
\begin{minipage}{0.45\textwidth}
  \caption[short figure description]{Time-averaged radial energy spectra $\widehat{E}(\nkk)$ as a function of $k = |\mathbf{k}|$ for different forcing $ |\Gamma_2/\Gamma_0|$. Dashed lines correspond to the condensate state. Characteristic scaling regimes are indicated. The gray region marks the energy input. The quantities are rescaled by $K=36.5$, the center of the region of energy input, and $|\widehat{E}^f_{\rm max}|$, the maximum of the energy spectrum in  this region. 
\label{fig3} 
}
\end{minipage}
\end{center}
\end{figure}

We consider regimes characterized by different behaviors of energy dissipation at large scales. This can be quantified by the radial energy spectrum of the kinetic energy $\widehat{E}(\nkk,t)= \frac{1}{2} \int _{\nkkp=\nkk}\uuh(\nkkp,t)^2 \dd\nkkp$ with $\uuh(\kk,t) = \int \uu(\xx,t) \exp^{-i\kk\cdot \xx} \dd\xx$ the Fourier transform of $\uu(\xx,t)$, see Section~\ref{sec:4}. The total energy, $E(t)= \int \widehat{E}(|\mathbf{k}|,t) \dd\nkk$ is fluctuating and eventually reaches a statistically steady state. Within this regime we consider the time-averaged spatial mean energy $\langle E \rangle_\Omega= \frac{1}{T} \frac{1}{|\Omega|} \int_{t_0}^{t_0+T} E(t) \dd t$, where $\Omega$ is the simulation domain, $t_0$ the time of statistical equalibration and $T$ the considered time interval. The results are shown in Figure \ref{fig2} as a function of the ratio $-\Gamma_2/\Gamma_0$. For $|\Gamma_2/\Gamma_0|\leq 2.4$, the values of $\langle E \rangle_\Omega$ are small and increase slightly with $|\Gamma_2/\Gamma_0|$. A sharp transition occurs at a critical value $|\Gamma_2/\Gamma_0|\approx 2.4$ where the flow begins to condensate and the mean energy jumps by an order of magnitude and increases more rapidly by further increasing $|\Gamma_2/\Gamma_0|\geq 2.4$. Below the transition, the corresponding flow fields are characterized by microvortices as shown in Figure \ref{fig1} a) and b). The distance of the microvortices is correlated to the unstable modes in the effective viscosity \cite{Slomkaetal_EPJSP_2015}. Above the critical value, the condensate is characterized by two counterrotating vortices, see Figure \ref{fig1} c). The size of these vortices depends on the considered domain size. This reproduces the analysis reported in \cite{Linkmannetal_PRL_2019}, where the presence of a condensate is associated with an inverse energy transfer. Indeed, analyzing this behavior in terms of energy spectra reveals the analogy with inertial turbulence for large values of $|\Gamma_2/\Gamma_0|$, see Figure \ref{fig3}. The time-averaged radial energy spectrum $\widehat{E}(\nkk) = \frac{1}{T} \int_{t_0}^{t_0 + T} \widehat{E}(\nkk,t) dt$ shows Kolmogorov scaling, $\widehat{E}(\nkk) \sim  \nkk^{-5/3}$ for these values. However, below the critical value, the spectra follow power laws, with activity-dependent non-universal scaling laws where the exponents are in the range set by energy equipartition, with $\widehat{E}(\nkk) \sim \nkk$ for low activities and $\widehat{E}(\nkk) \sim \text{const.}$ close to the critical value. 

\begin{figure*}[htb]
  \noindent
  \begin{tabular}{ll}
  \begin{picture}(0.45 \textwidth, 0.35 \textwidth)
     \put(0,0){\includegraphics*[width = 0.45 \textwidth ]{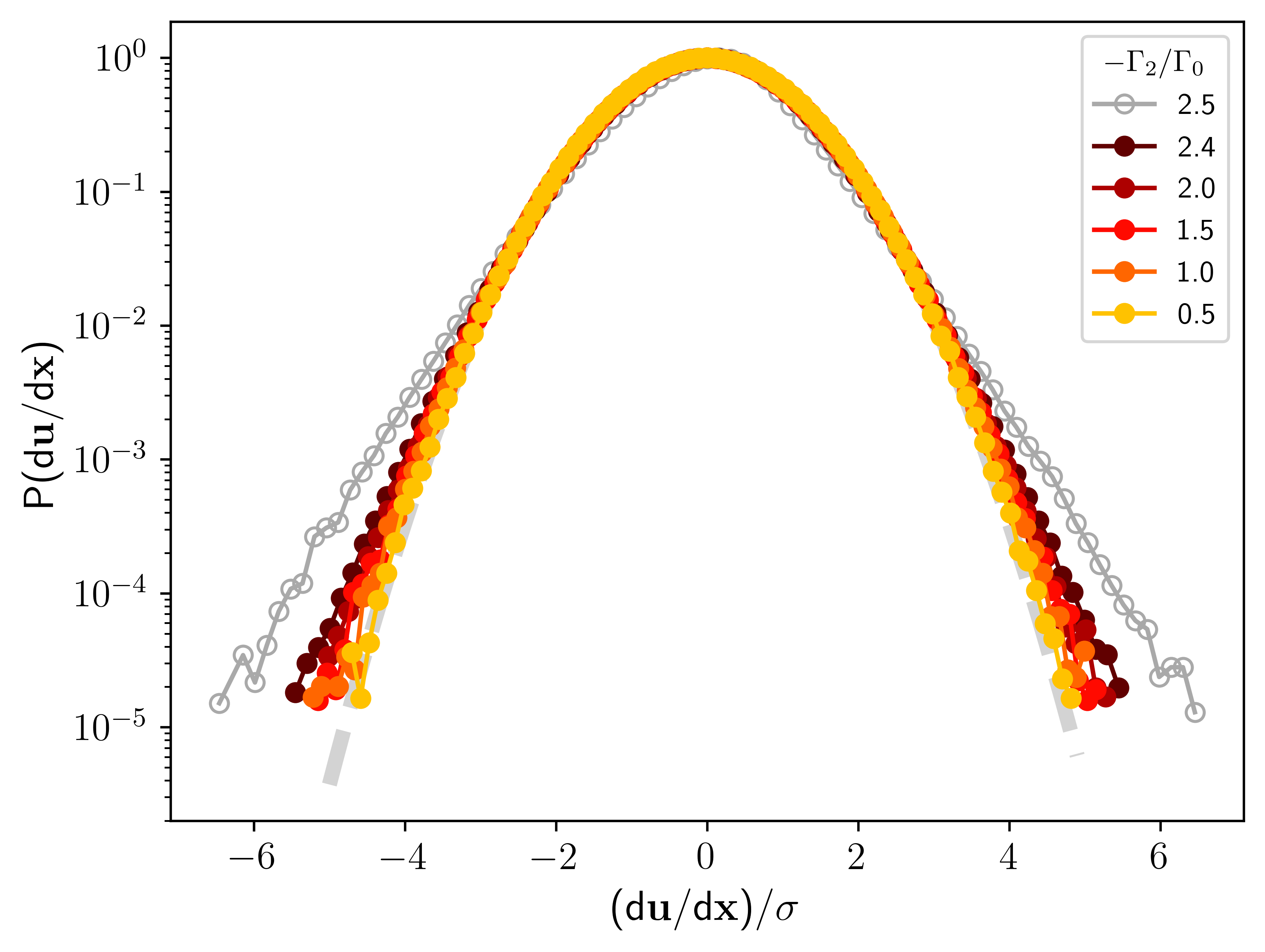}}
     \put(4,8){a)}
    \end{picture}
    &  \begin{picture}(0.45 \textwidth, 0.35 \textwidth)
     \put(0,0){\includegraphics*[width = 0.45 \textwidth]{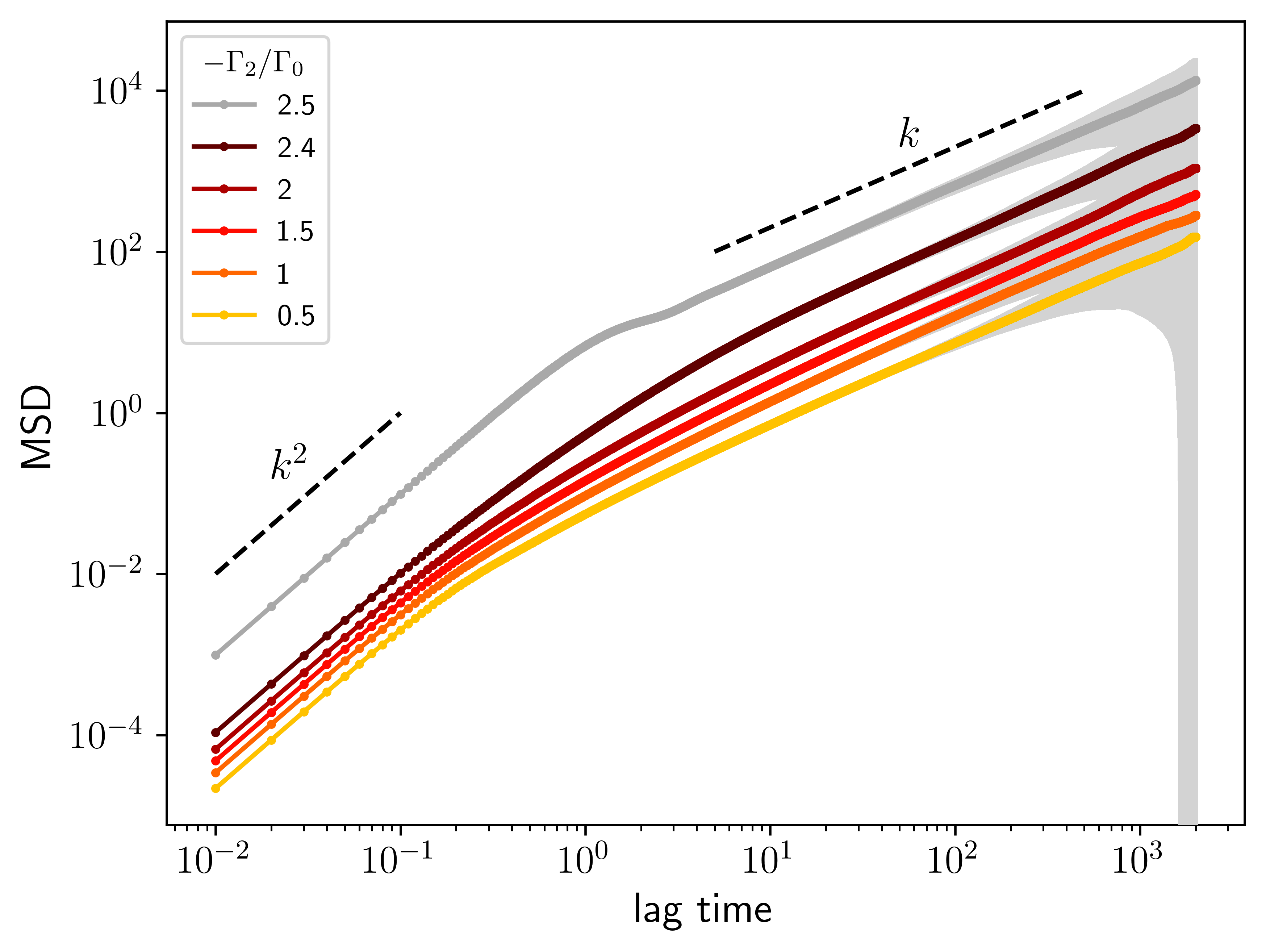}}
     \put(4,8){c)}
     \end{picture}\\
    &   \\
    \begin{picture}(0.45 \textwidth, 0.18 \textwidth)
     \put(0,0){\includegraphics*[width = 0.45 \textwidth ]{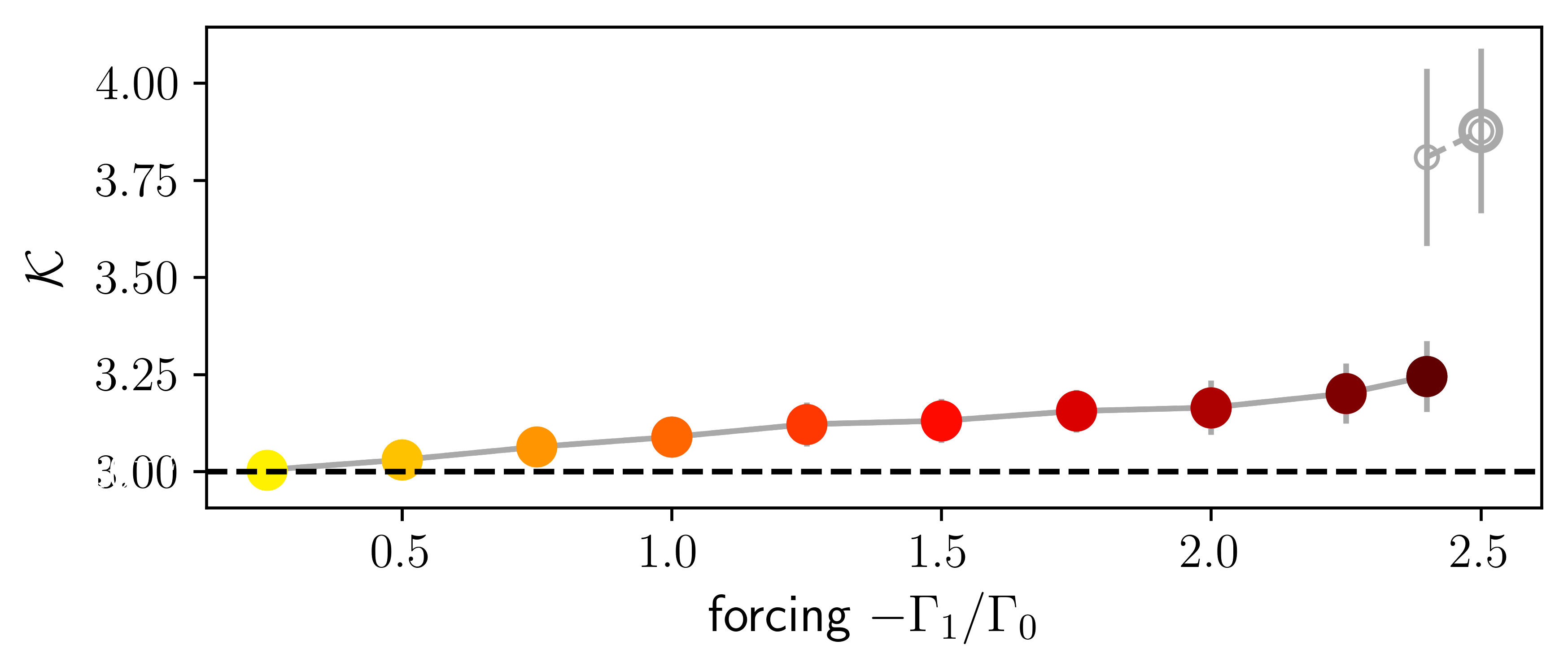}}
\put(4,8){b)}
     \end{picture}
    &\begin{picture}(0.45 \textwidth, 0.18 \textwidth)
     \put(0,0){\includegraphics*[width = 0.45 \textwidth ]{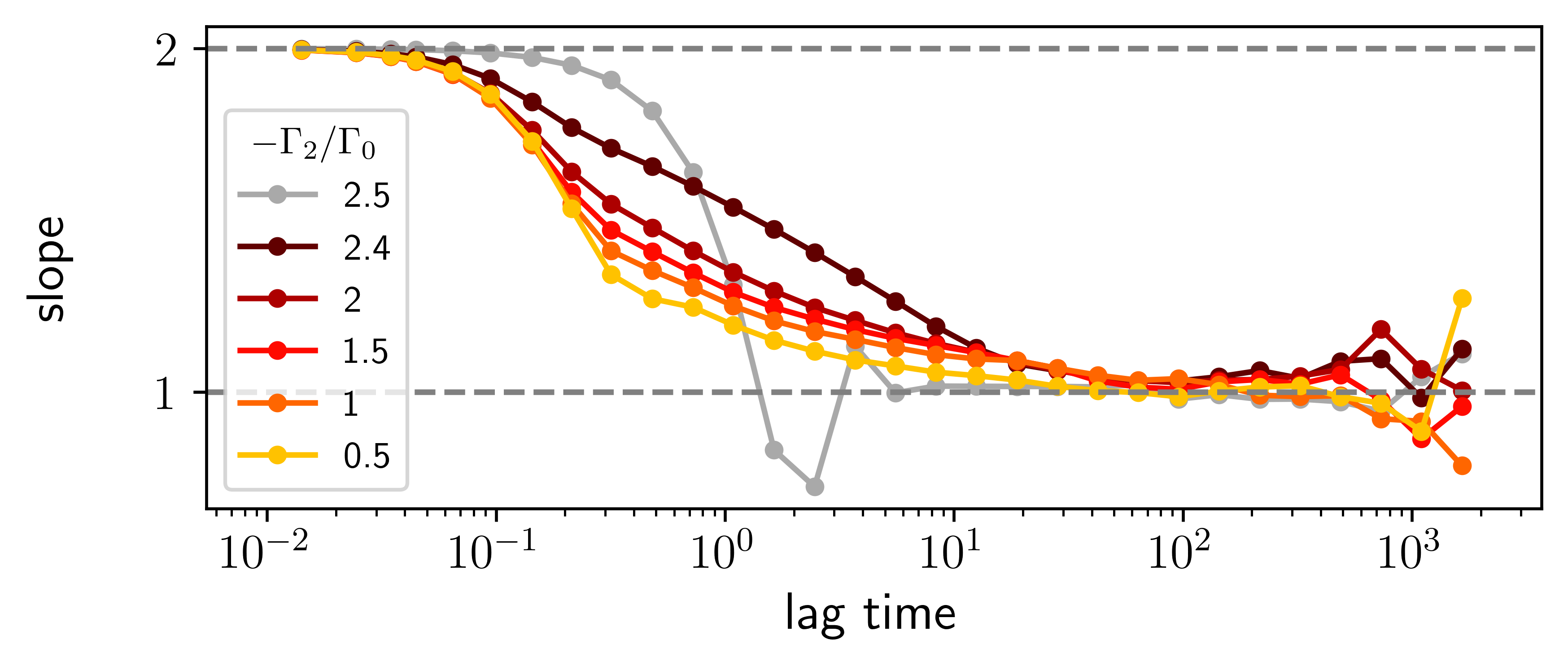}}
     \put(4,8){d)}
     \end{picture}
     \end{tabular}
  \begin{center}
\begin{minipage}{0.95\textwidth}
  \caption[short figure description]{ a) Infinitesimal (longitudinal) velocity increment evaluated by the local gradient of the velocity field $\mathbf{u}(\mathbf{x},t)$. Normalized such that the standard deviation is 1 and compared with a Gaussian distribution (dashed line). b) Kurtosis $\mathcal{K}$ for different forcing. c) Mean square displacement (MSD). The grey shading indicates the standard deviation resulting from the averaging of MSD of all trajectories. d) The slope of MSD shown in c). All considered for different forcing $|\Gamma_2/\Gamma_0|$. 
\label{fig4}
} 
\end{minipage}
\end{center}
\end{figure*}
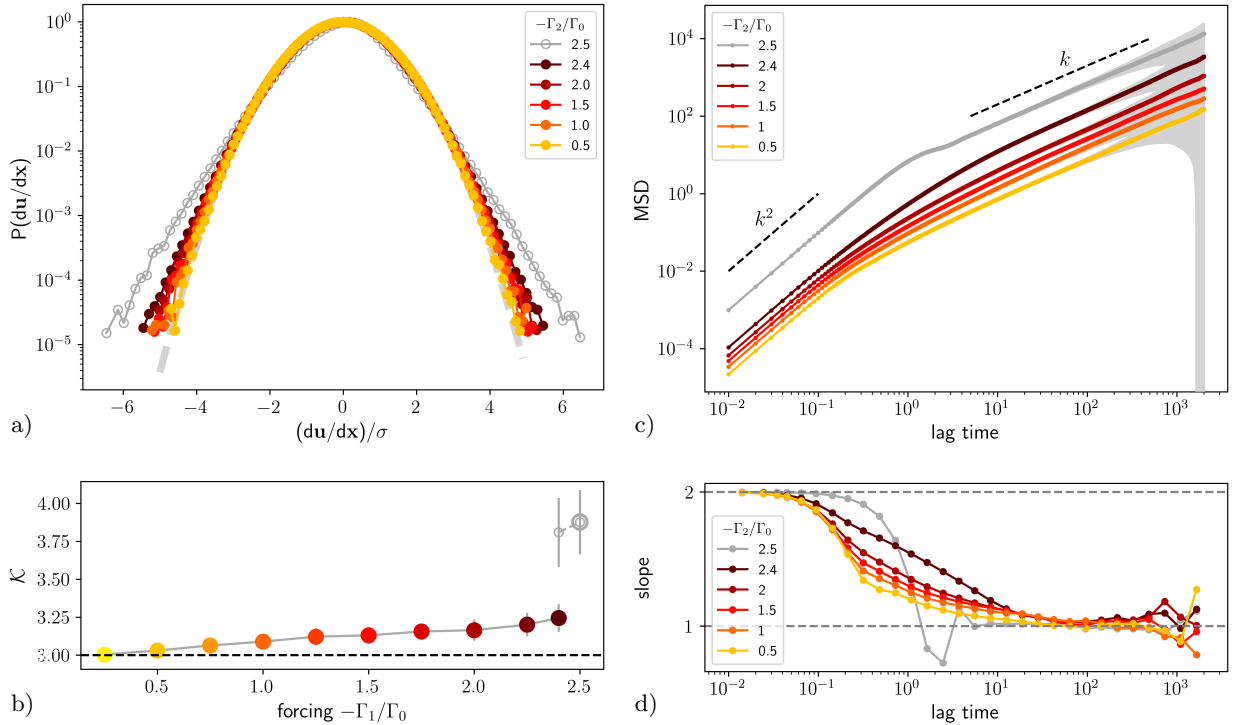

The transition between active turbulence states without and with characteristics of inertial turbulence is further supported by considering the distribution of (longitudinal) velocity increments, see Section \ref{sec:4}. A significant deviation from a Gaussian distribution is a signature of fully developed inertial turbulence where fat tails suggest the velocity increments are intermittent \cite{Boffettaetal_ARFM_2012}. The distributions, together with a Gaussian distribution for comparison, are shown in Figure \ref{fig4} a) and lead to similar results as in \cite{Mukherjeeetal_NP_2023}. The deviation from a Gaussian distribution increases with increasing activity. It becomes distinctly non-Gaussian above some critical value, which can be quantified by the kurtosis $\mathcal{K}$, see Section \ref{sec:4}, shown in Figure \ref{fig4} b). Starting with $\mathcal{K} = 3$, the kurtosis slightly increases until the critical value, where a sharp transition occurs, and $\mathcal{K} \gg 3$ for larger activities. A similar non-Gaussian behaviour was also observed in velocity gradient statistics for bacterial suspensions \cite{Ilkanaivetal_PRL_2017}. Besides these Eulerian measures, Figure \ref{fig1} d) - f) show trajectories of a tracer particle within the flows depicted in Figure \ref{fig1} a) - c). The spatial scale is adapted to fit the trajectories independent of the periodic boundary conditions. The qualitative differences are already visible. As in \cite{Mukherjeeetal_PRL_2021}, we compute the mean square displacement (MSD) where we consider the ensemble average over 100 particles. The results are shown in Figure \ref{fig4} c). We observe a turnover from a ballistic regime $\Delta x^2 \sim t^2$ to a diffusive regime $\Delta x^2 \sim t$. Above the critical value, this turnover is more or less sharp, whereas below the critical value, it is smooth with an extended intermediate anomalous diffusive regime with $\Delta x^2 \sim t^\xi$ with $1 < \xi < 2$. The identified slopes $\xi$, as they evolve over time, are shown in Figure \ref{fig4} d). Below the critical value, the intermediate anomalous diffusive regime increases with activity. However, above the critical value the qualitative behavior changes due to the formation of condensates strongly influencing the trajectories. In \cite{Mukherjeeetal_PRL_2021} this superdiffusive behavior is related to experiments for dense bacterial suspensions \cite{Wuetal_PRL_2000,Arieletal_NC_2015}. As in \cite{Mukherjeeetal_PRL_2021} we refrain from more detailed analysis due to a lack of high-precision data. 

These different measurements not only confirm known characteristics of inertia turbulence for active turbulence if the considered forcing is above some critical threshold, but they also confirm the existence of a threshold and provide different approaches to measure it quantitatively. The sharp transition for the time-averaged spatial mean energy in Figure \ref{fig2}, the change in scaling of the time-averaged radial energy spectra to Kolmogorov scaling in Figure \ref{fig3}, the strong deviation from a Gaussian distribution of the velocity increments in Figure \ref{fig4} a), quantified by a substantial increase of the kurtosis in Figure \ref{fig4} b), and the change in transition from a ballistic to a diffusive regime in the MSD in Figure \ref{fig4} c), quantified by the change in slope in Figure \ref{fig4} d), all occur at $|\Gamma_2/\Gamma_0| = 2.4$. Below this threshold, all considered measures only slightly increase with forcing.

\begin{figure}[htb]
  \noindent
  \begin{tabular}{l}
    \includegraphics*[width = 0.45 \textwidth ]{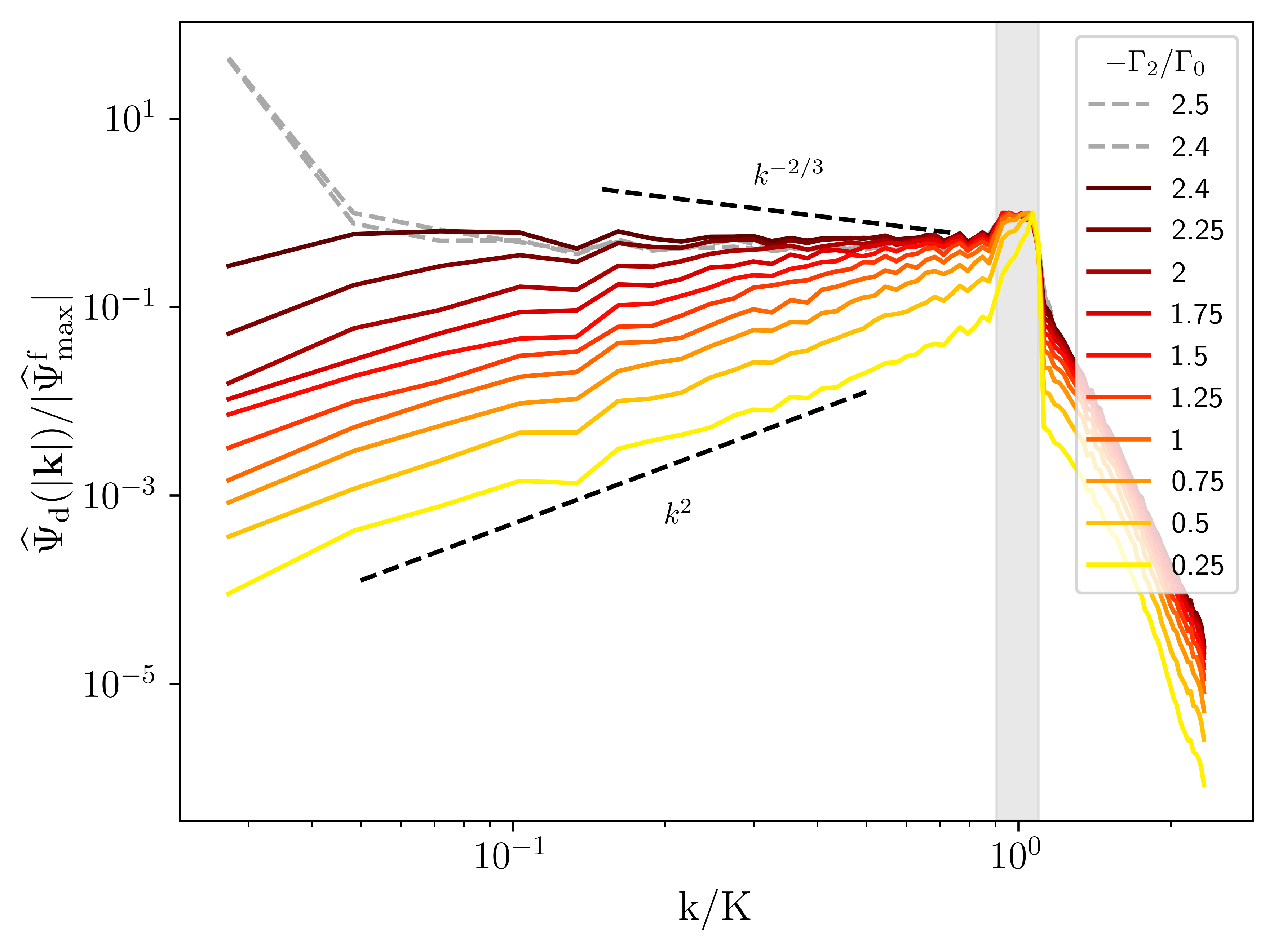}
  \end{tabular}
  \begin{center}
\begin{minipage}{0.45\textwidth}
  \caption[short figure description]{
    Radial spectra of autocovariance function $\widehat{\Psi}_{\rm d}(\nkk)$ as a function of $k = |\mathbf{k}|$ for different forcing $|\Gamma_2/\Gamma_0|$. Dashed lines correspond to the condensate state. Characteristic scaling regimes are indicated. The gray region marks the energy input. The quantities are rescaled by $K=36.5$, the center of the region of energy input and $\widehat{\Psi}^f_{\rm max}$, the maximum of $\widehat{\Psi}_{\rm d}(\nkk)$ in this region.
\label{fig5} 
}
\end{minipage}
\end{center}
\end{figure}

We add a new measure to support the described transition and to characterize the system below the critical forcing. We test for hyperuniformity of the spatial order of the microvortices, which characterize the statistically steady states. We consider two different approaches; see Section \ref{sec:4}. The first extracts the extrema of the vorticity field $\ww(\mathbf{x}, t)$ and analyses the resulting point configurations $\{\mathbf{r}_j\}$, see SI Figure 1 a) - c) for corresponding points to Figure \ref{fig1} a) - c). SI Figure 1 d) - f) show the corresponding scattering patterns with a circular region around the origin where there is almost no scattering for low activities. This is a first signature for long-range correlations, which may correspond to a hyperuniform character. More quantitatively, SI Figure 2 a) and b) show the structure factor $\widehat{S}(\nkk)$ and the value at largest length scale $\widehat{S}( \nkk=\nkk_{\rm min})$ as a function of forcing strength. While the last increases with forcing and shows a jump at the critical value $|\Gamma_2/\Gamma_0| = 2.4$, the data in SI Figure 2 a) is too noisy to analyze the behavior for $|\mathbf{k}| \to 0$. The second approach uses an extension of the concept of hyperuniformity to scalar fields \cite{Maetal_JAP_2017,Tor16,Tor18,salvalaglio2020hyperuniform} and analyses the vorticity field $\ww(\mathbf{x}, t)$ directly. Defining the autocovariance function 
\begin{equation}\label{eq:psi}
\psi(\mathbf{r})=\langle [\ww(\mathbf{x}_1,t)-\langle \ww(\mathbf{x}_1,t)\rangle ] 
[\ww(\mathbf{x}_2,t)-\langle \ww(\mathbf{x}_2,t)\rangle ]\rangle
\end{equation}
with $\mathbf{r}=\mathbf{x}_2-\mathbf{x}_1$ and $\langle\ \cdot \ \rangle$ the ensemble average over time instances, we consider its spectral density $\widehat{\psi}(\mathbf{k}) = \widehat{\psi}_d(|\mathbf{k}|) 2 \pi |\mathbf{k}|$, with the radial spectrum $\widehat{\psi}_d(|\mathbf{k}|)$, which is shown in Figure \ref{fig5}. Hyperuniform characters can be inspected by considering the scaling and the values for small wavenumbers (see, e.g., \cite{Maetal_JAP_2017}). Again we observe a strong dependency on the forcing with a qualitative change in the behaviour at the critical value $|\Gamma_2/\Gamma_0| = 2.4$. A transition towards "anti-hyperuniformity" can be seen. Below, the critical value characteristics of hyperuniformity can be identified. Scaling with $|\mathbf{k}|^\alpha$ and $\alpha \sim 2$ can be observed for forcing significantly below the critical value. The measured values for the smallest wavevector increase from $10^{-4}$ to $10^{-1}$ with increasing forcing. These results indicate Type I hyperuniformity \cite{Tor18} for low forcing and a possible transition to non-hyperuniformity close to the critical value. 

By invoking ergodicity, so implying statistically homogeneous $\psi(\mathbf{r})$, and assuming an infinite system, the spectral density $\widehat{\psi}(\mathbf{k})$ corresponds to $|\widehat{\omega}(\mathbf{k})|^2$ 
with $\widehat{\omega}(\mathbf{k}) = \frac{1}{T} \int_{t_0}^{t_0 + T} \widehat{\omega}(\mathbf{k},t) dt $ and $ \widehat{\omega}(\mathbf{k},t) $ the Fourier transform of the vorticity $\omega(\mathbf{x},t)$ 
\cite{Torquato1999}. This allows for a direct relation between the energy spectra in Figure \ref{fig3} and the autocovariance spectra in Figure \ref{fig5}, resulting from \begin{equation} \label{eq:scaling}
    \widehat{E}(\nkk) = \frac{|\wwh(\mathbf{k})|^2}{\nkk^2}=\frac{2 \pi \widehat{\Psi}_{\rm d}(\nkk)}{\nkk},
\end{equation}
where the first relation is an established connection in fluid mechanics between the energy $E(\mathbf{u})$ and the enstrophy ${\cal{E}}(\omega) = \int |\omega(\mathbf{x})|^2 d \mathbf{x}$, see \cite{Alertetal_ARCMP_2022}. The indicated scaling exponents in Figure \ref{fig5} support this relation and allow for a broader discussion.

\section{Discussion}
\label{sec:3}

\begin{figure}[t]
  \noindent
  \begin{tabular}{l}
  \begin{picture}(0.45 \textwidth,0.35 \textwidth)
    \put(0,0){\includegraphics[width = 0.45 \textwidth ]{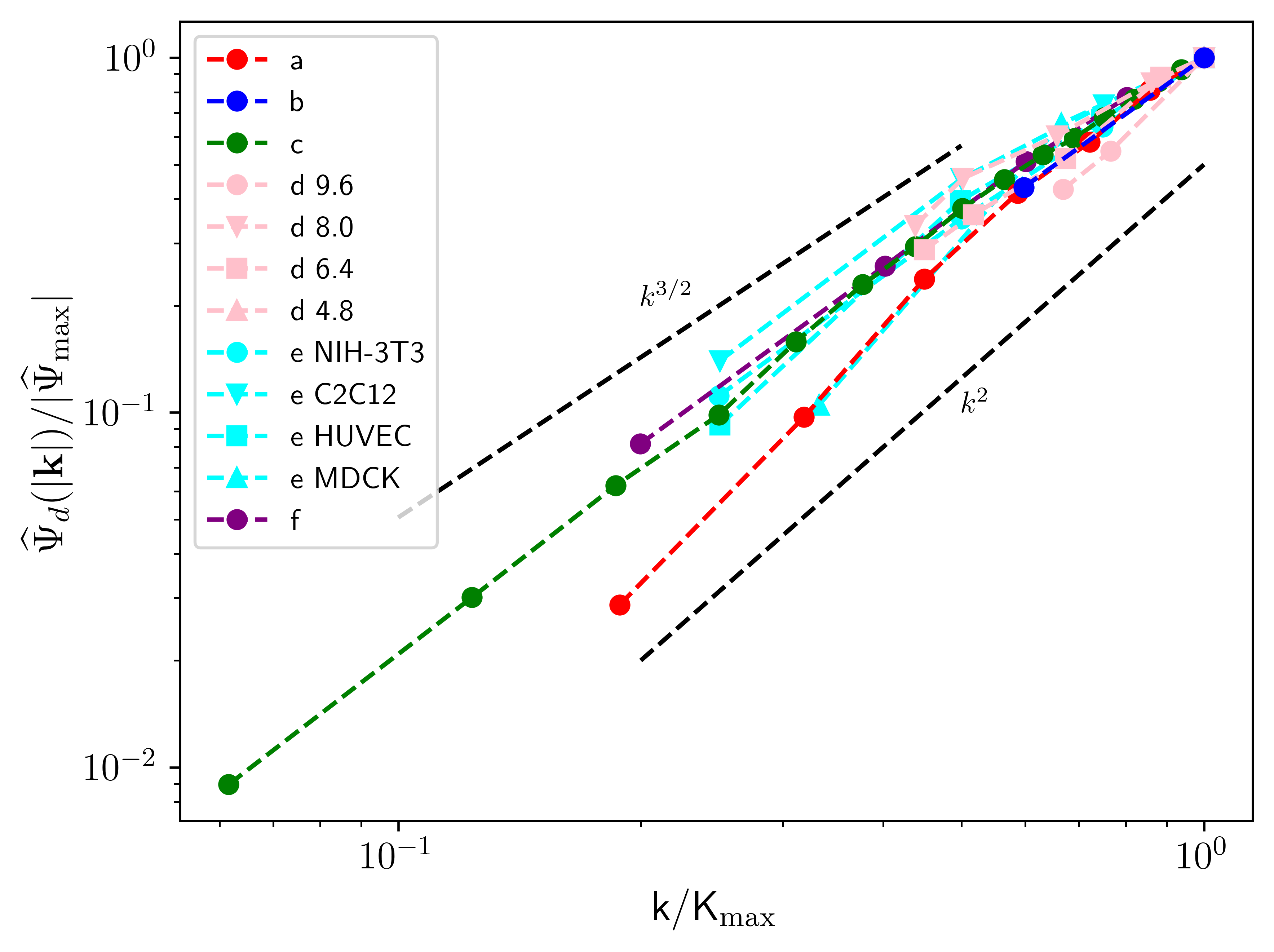}}
    \put(4,8){a)}
    \end{picture}
    \\
     \begin{picture}(0.45 \textwidth,0.23 \textwidth)
    \put(0,0){\includegraphics[width = 0.45 \textwidth ]{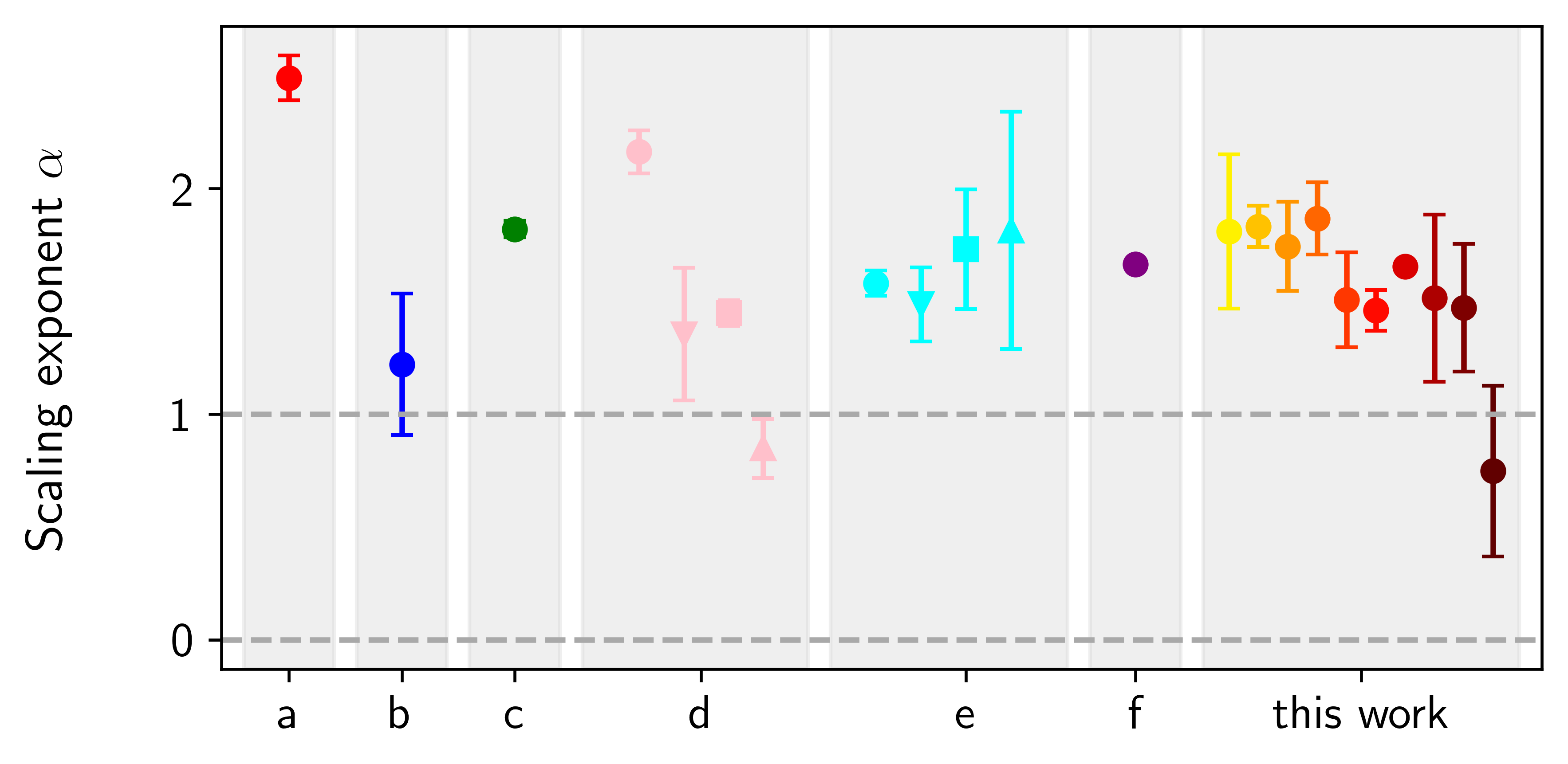}} 
    \put(4,0){b)}
    \end{picture}
  \end{tabular}
  \begin{center}
\begin{minipage}{0.45\textwidth}
  \caption[short figure description]{
    a) Radial spectra of autocovariance function $\widehat{\Psi}_{\rm d}(\nkk)$ as a function of $k = |\mathbf{k}|$ for various experimental systems taken from \cite{Alertetal_ARCMP_2022} (Figure 1, corresponding to a) bacterial suspension (B. subtilis) \cite{Wensinketal_PNAS_2012}, b) sperm suspension \cite{Creppyetal_PRE_2015}, c) self-propelled Janus particles \cite{Nishiguchietal_PRE_2015}, d) bacterial suspension (E. coli) \cite{Pengetal_SA_2021}, the numbers correspond to selected volume fractions, e) tissue cell monolayers \cite{Linetal_CP_2021} for different tissues (NIH-3T3, mouse embryonic fibroblast cell line; C2C12, mouse myoblast cell line; HUVEC, human umbilical vein endothelial cells; MDCK, Madin-Darby Canine Kidney cells) and f) microtubulin-kinesin suspensions \cite{Martinez-Pratetal_PRX_2021}). Data points are rescaled and normalized by $K_{\rm max}$, and $\widehat{\Psi}_{\rm max}$. \eqref{eq:scaling} is used to transform the energy spectra into the autocovariance function. Characteristic scaling regimes are indicated. b) Scaling exponent $\alpha$ obtained from the three lowest values of $k$ for each curve in a) and Figure \ref{fig5}, together with characteristic values for different types of hyperuniformity. Same labels as in a).  
\label{fig6} 
}
\end{minipage}
\end{center}
\end{figure}

Our results are not restricted to the considered GNS equations. The established connection between scaling properties of the autocovariance function and the energy spectra also allows testing various experimental systems for hyperuniformity. Figure \ref{fig6} a) summarizes results reported in \cite{Alertetal_ARCMP_2022}. These data contain polar as well as nematic systems and with the exception of \cite{Pengetal_SA_2021}, all systems can be considered as (quasi) two-dimensional. None of these studies focuses on low active forcing, and most are concerned with intermediate scales and the comparison with inertial turbulence. However, rescaling the measured energy spectra in these experiments, applying the scaling transform in \eqref{eq:scaling} and focusing on $|\mathbf{k}| \to 0$ allows to extract the scaling exponent $\alpha$, from which characteristics of hyperuniformity can be obtained. The results are shown in Figure \ref{fig6} b) together with our simulation data. Using the classification of hyperuniformity in \cite{Tor18} ($\alpha > 1$ type I, $\alpha = 1$ type II, $0 < \alpha < 1$ type III, $\alpha = 0$ non-hyperuniformity, $\alpha < 0$ anti-hyperuniformity) almost all systems indicate type I hyperuniformity. Only the lowest volume fractions of the bacterial suspension (E. coli) \cite{Pengetal_SA_2021} and the largest forcing before condensation in our simulations indicate type 3 hyperuniformity. Further increasing the forcing in the simulations leads to non- and anti-hyperuniformity. This conclusion relies only on the scaling behaviour. Considering in addition the values of $\widehat{\Psi}_d(|\mathbf{k}|)/ |\widehat{\Psi}_{\rm max}|$ for the smallest values of $|\mathbf{k}|$ available in Figure \ref{fig6} a), which gives a finite size estimate of the hyperuniformity metric $H = \lim_{|\mathbf{k}| \to 0} \widehat{\Psi}_d(|\mathbf{k}|) / |\widehat{\Psi}_{\rm max}|$, see \cite{Maetal_JAP_2017}, this conclusion is less clear for most of the experimental data. While ideally $H = 0$ for hyperuniform systems, in the literature it is commonly accepted the nomenclature of near hyperuniformity for $H\leq 10^{-2}$ and effective hyperuniformity for $H\leq 10^{-4}$ \cite{Torquato2006,Kim2018}. The first could be confirmed only for the self-propelled Janus particles in \cite{Nishiguchietal_PRE_2015}. All others ask for measurements on larger systems. However, the simulation results, see also the selected simulations on larger domains in SI Figure 3, are clearly within the classification of nearly hyperuniform systems, at least for forcing significantly below the critical value, and the classification as type I hyperuniformity holds.

Besides the theoretical interest in the context of non-equilibrium physics, the connection between hyperuniformity and active turbulence also allows for speculations from a biological perspective. 
Combining general properties of disordered hyperuniform systems and encountered Lagrangian measures of Levy walks and non-universal diffusion allows for the following interpretation: The systems invest energy on a low level to maintain a non-equilibrium hyperuniform state, characterized by isotropic properties and robustness against defects \cite{Tor18} which are preferential properties of a resting state. Only if appropriate, the systems invest larger activities, relaxing the hyperuniform state, leading to different classifications and types and states with enhanced superdiffusive properties, which are optimal strategies for evasion and foraging \cite{Humphriesetal_PNAS_2012}. 
Similar strategies have been identified for a single bacterium \cite{Korobkovaetal_Nature_2004} and T cells \cite{Harrisetal_Nature_2012}, which switch between diffusive and superdiffusive motility in search for dense and sparsely distributed targets, respectively. Such behaviour has also been modeled \cite{Estrada-Rodriguezetal_JNS_2022}, and even robot swarms use such strategies, see \cite{Duncanetal_BB_2022} and the references therein. However, if this plausible-seeming strategy, which could be drawn from our simulation results, can explain why nature allows for such complex systems, has to be further explored.

\section{Materials and methods}
\label{sec:4}

\eqref{eq:GNS} is transformed into a vorticity formulation using $\omega \mathbf{e}_3 = \nabla \times \mathbf{u}$ with $\omega(\mathbf{x},t)$ the vorticity and $\mathbf{e}_3(\mathbf{x})$ the unit vector perpendicular to the considered two-dimensional space. The resulting equation reads
\begin{align}
\partial_t \ww + \uu \cdot \nabla \ww = \nueff \Lap \ww. \label{eq:GNSV}
\end{align}
We consider the piecewise-constant approximation
\begin{align}
\nueff(|\mathbf{k}|)=
\begin{cases}
\GO>0 & \text{for } |\mathbf{k}| < k_0\\
\GT<0 & \text{for } k_0 \leq |\mathbf{k}| \leq k_1 \\
\GF>0 & \text{for } k_1 < |\mathbf{k}|
\end{cases}
\end{align}
as introduced and justified in \cite{Linkmannetal_PRL_2019} and solve \eqref{eq:GNSV} 
using a pseudospectral method with a third-order TVD (Total Variation Diminishing) scheme for time discretization and a periodic domain of size $[-\pi,\pi] \times [-\pi,\pi]$ which is discretized by $2^{8}$ points in each direction. For selected simulations we also increase the domain size to demonstrate the persistence of the scaling on larger scales, see SI Figure 3. The domain size influences the onset of the deviation above the threshold. However, the results concerning the threshold and characteristics of hyperuniformity are not affected.

In our simulations, fixed values of $\GO=1.1\times10^{-3}$, $\GF=1.1\times10^{-2}$, $k_0 = 33$ and $k_1 = 40$ are chosen, while $\GT$ is varied. The effective driving of the fluid is defined as the ratio of the forcing viscosity to the viscosity for long wavelengths, $-\GT/\GO$, and the forced wave numbers are confined to $[k_0, k_1]$. For all simulations, the initial data are Gaussian-distributed random velocity fields. All simulations run until a statistically steady state is reached, and data are analysed only within this state. 

The statistically steady states are characterized by microvortices, which don't or do form a condensate depending on the activity strength. We consider two different approaches to analyse the spatial order of these microvortices. The first extracts the extrema of the vorticity field $\ww(\mathbf{x},t)$ and analyses the resulting point configurations $\{\mathbf{r}_j\}$, see SI Figure 1 a) - c) for corresponding points to Figure \ref{fig1} a) - c). Hyperuniformity in these point configurations can be characterized by computing $\widehat{S}(|\mathbf{k}|) \sim |\mathbf{k}|^\alpha$ as $|\mathbf{k}| \to 0$ or $\sigma^2(R) \sim R^\beta$ as $R \to \infty$ and determining $\alpha$ and $\beta$. Both criteria are connected by defining the density 
\begin{equation}
    \widehat{\rho}(\mathbf{k},\{\mathbf{r}_j\})=\sum_{j=1}^N\exp({\rm i}\mathbf{k}\cdot\mathbf{r}_j),
\end{equation}
with $\mathbf{r}_j$ the position of point $j$ and $N$ the number of points. The structure factor now reads 
\begin{equation}
    \widehat{S}(\mathbf{k},\{\mathbf{r}_j\})=\frac{|\widehat{\rho}(\mathbf{k},\{\mathbf{r}_j\})|^2}{N}=1+\frac{2}{N}\widehat{C}(\mathbf{k},\{\mathbf{r}_j\}),
\end{equation}
with $\widehat{C}(\mathbf{k},\{\mathbf{r}_j\})$ the real quantity
\begin{equation}
    \widehat{C}(\mathbf{k},\{\mathbf{r}_j\})=\sum_{j=1}^{N-1} \sum_{m=j+1}^N\cos(\mathbf{k}\cdot(\mathbf{r}_j-\mathbf{r}_m)).
\end{equation}
Results are shown in SI Figure 2. The second approach uses an extension of the concept of hyperuniformity to scalar fields \cite{Maetal_JAP_2017,Tor16,Tor18,salvalaglio2020hyperuniform} and analyses the vorticity field $\ww(\mathbf{x}, t)$ directly. Using the autocovariance function $\psi(\mathbf{r})$, see \eqref{eq:psi}, the field is hyperuniform if the radial spectral density $\widehat{\psi}_{\rm d}(|\mathbf{k}|)$, which is the radial average of the Fourier transform of $\psi(\mathbf{r})$, satisfies the condition
\begin{equation}
\widehat{\psi}_{\rm d}({|\mathbf{k}}|) \to 0 \quad \mbox{as} \quad |\mathbf{k}|\rightarrow 0
\end{equation}
and hyperuniform characters can be inspected by the associated scaling of $\widehat{\psi}_{\rm d}(|\mathbf{k}|)$ for small wavenumbers (see, e.g., \cite{Maetal_JAP_2017}). 
As the identification of extrema is associated with additional numerical error, and the definition of $\ww(\mathbf{x},t)$ inherently contains more information, we favor the second approach. 

%\textcolor{red}{Averages from the statistically steady states are computed using at least ? snapshots.}

The velocity increments are computed as $\delta_\vl \uu := \uu(\xx+\vl)-\uu(\xx)$ which is a vector and depends on the length and direction of \vl \cite{Mukherjeeetal_PRL_2021}. A longitudinal velocity increment (in velocity direction) is defined as $\delta_{\parallel,l} \uu := \delta_\vl \uu \cdot \frac{\uu}{|\uu|}$. Assuming Taylor hypothesis we obtain $\delta_\vl \uu := \uu(\xx+\vl)-\uu(\xx) \approx \vl \frac{\partial \uu}{\partial \xx} = \vl_i \partial_i u_j$ for $l \rightarrow 0$, which allows to define the longitudinal velocity increment as
\begin{align}
\delta_\parallel \uu = \hat{\uu}  \frac{\partial \uu}{\partial \xx} \hat{\uu}
  = \hat{u}_i  \partial_i u_j \hat{u}_j \quad \mathrm{with} \quad \hat{\uu}= \frac{\uu}{|\uu|}
\end{align}

Deviations from a Gaussian distribution, which are a signature of inertial turbulence \cite{LM05,LPE08} are quantified by the kurtosis $\mathcal{K}$, which is computed as the fourth moment of the velocity increment  
\begin{align}
\mathcal{K}=\frac{\langle  (\delta_\vl \uu)^4 \rangle}{\langle (\delta_\vl \uu)^2 \rangle^2} .
\end{align}
as in \cite{YD18}. \\

\appendix

\section{Point configurations and structure factor}

\begin{figure*}[htb]
  \noindent
  \begin{tabular}{ll}
    \includegraphics*[width = 0.24 \textwidth]{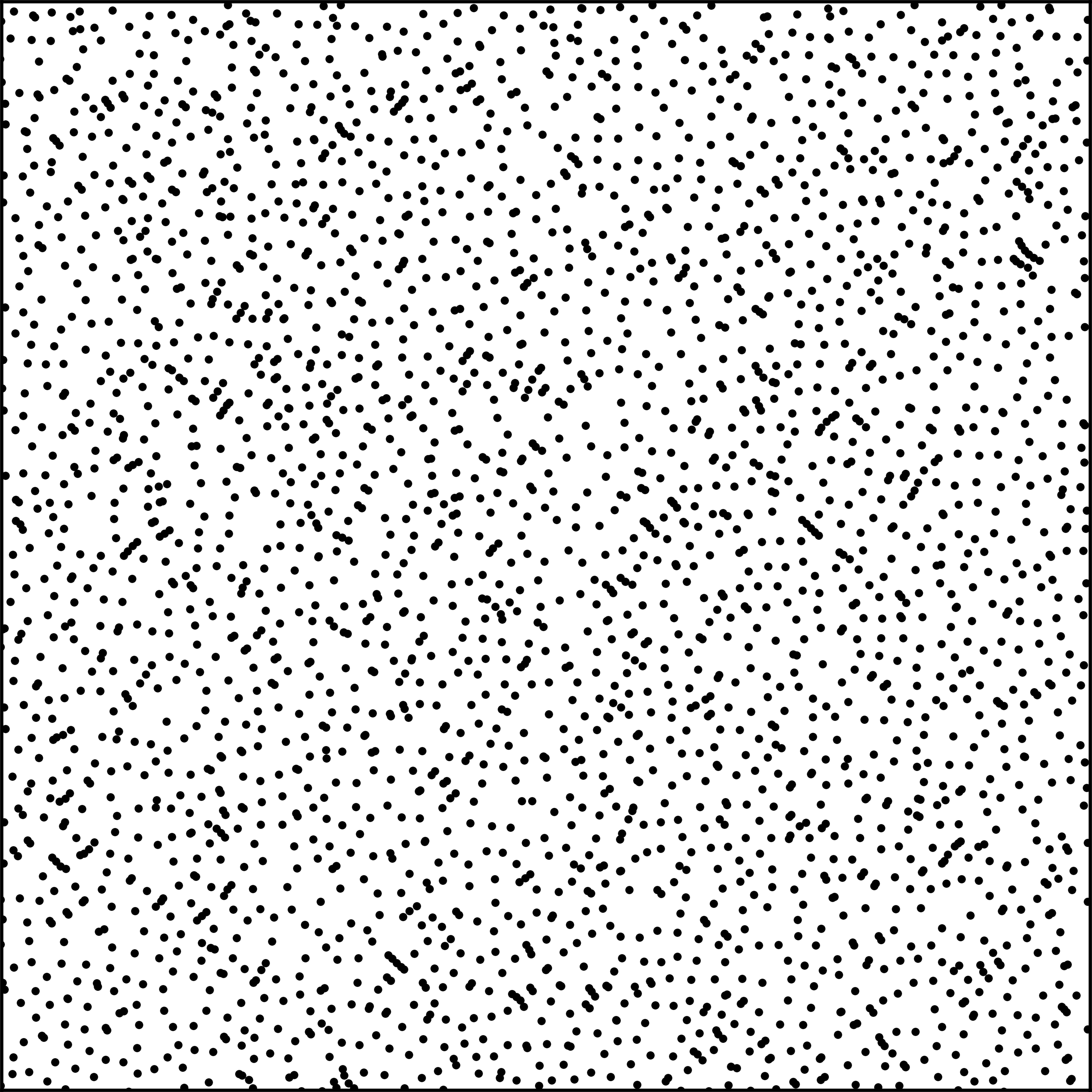}
    &\includegraphics*[width = 0.24 \textwidth]{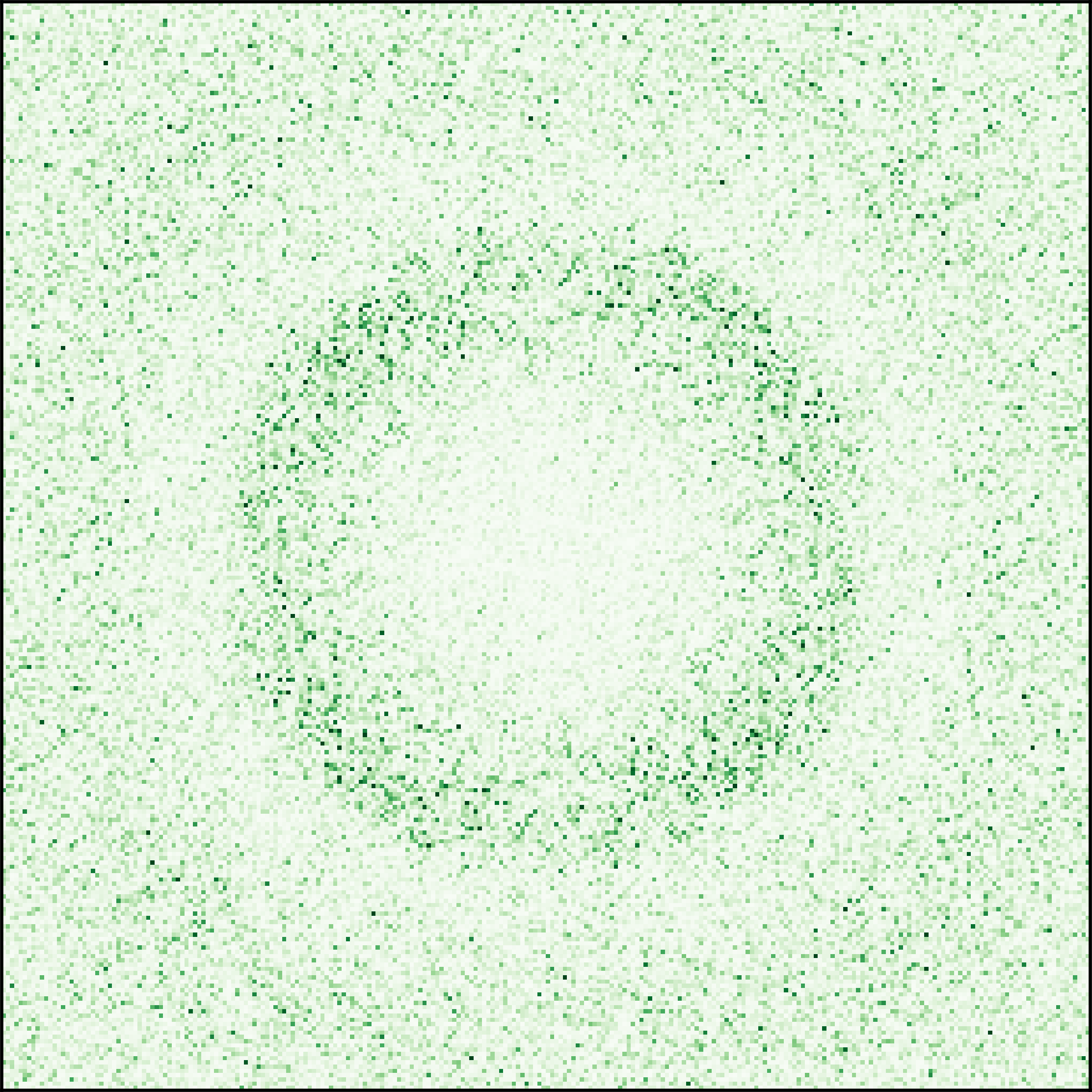}\\
    a) & d) \\
    \includegraphics*[width = 0.24 \textwidth ]{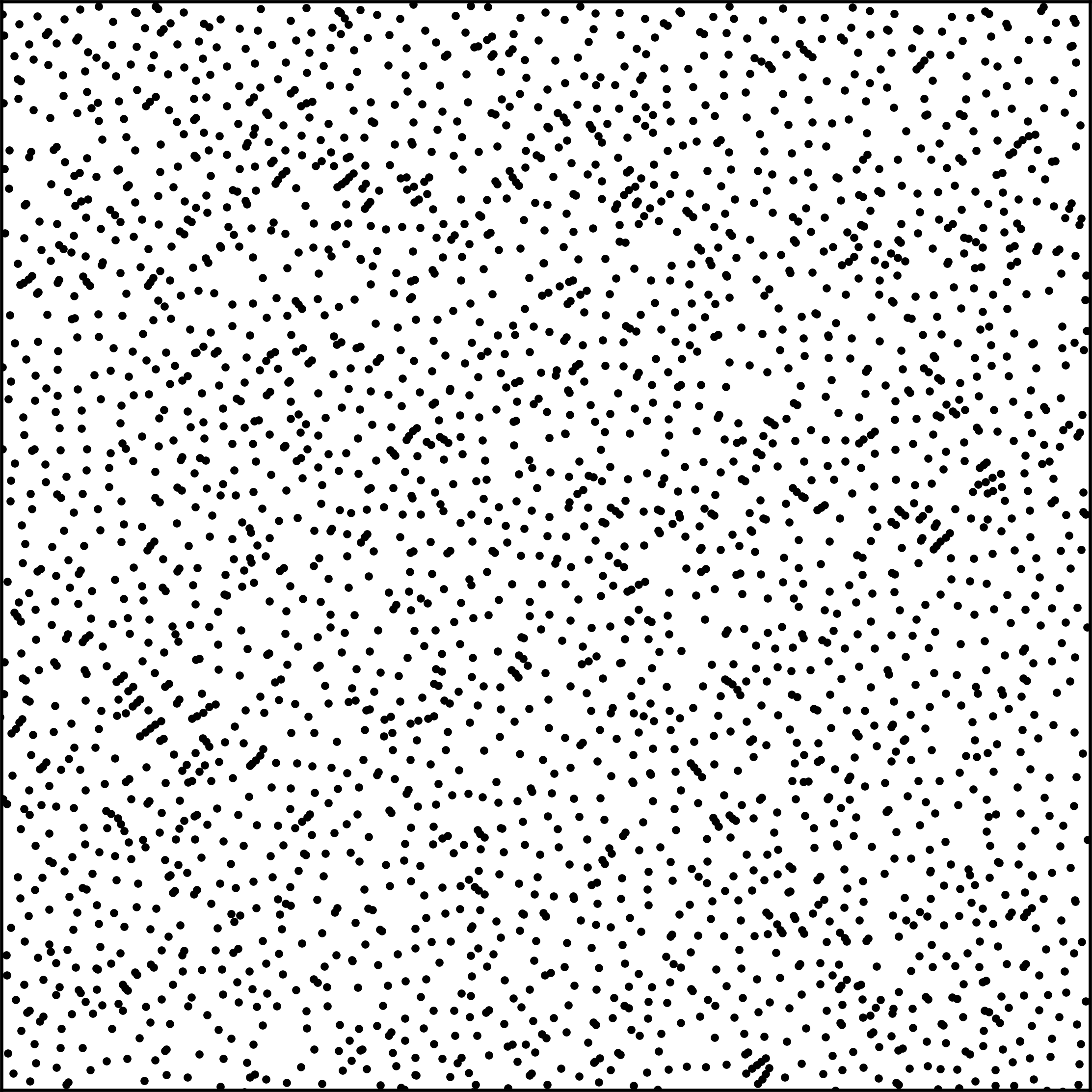} 
    &\includegraphics*[width = 0.24 \textwidth ]{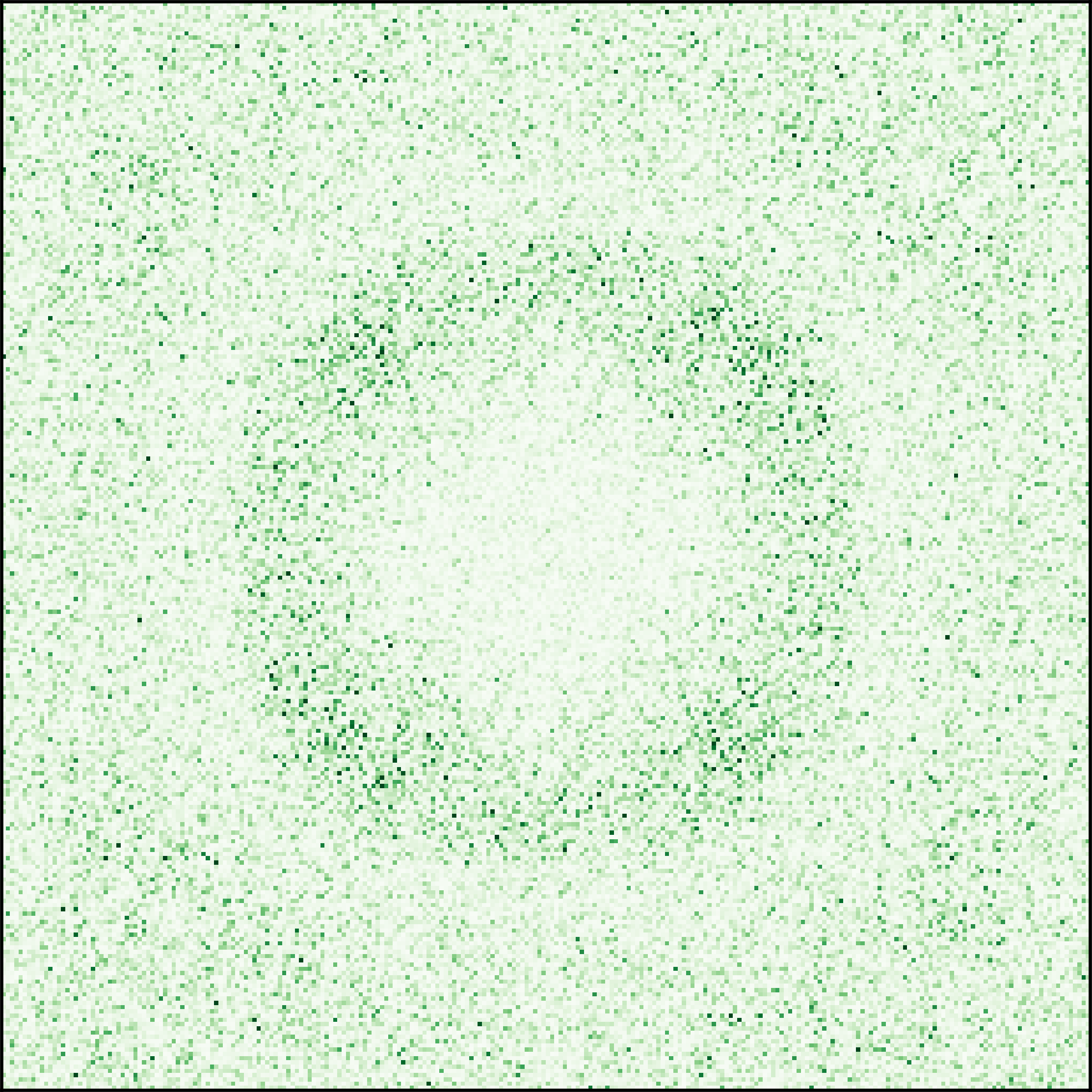}\\
    b) & e) \\
    \includegraphics*[width = 0.24 \textwidth]{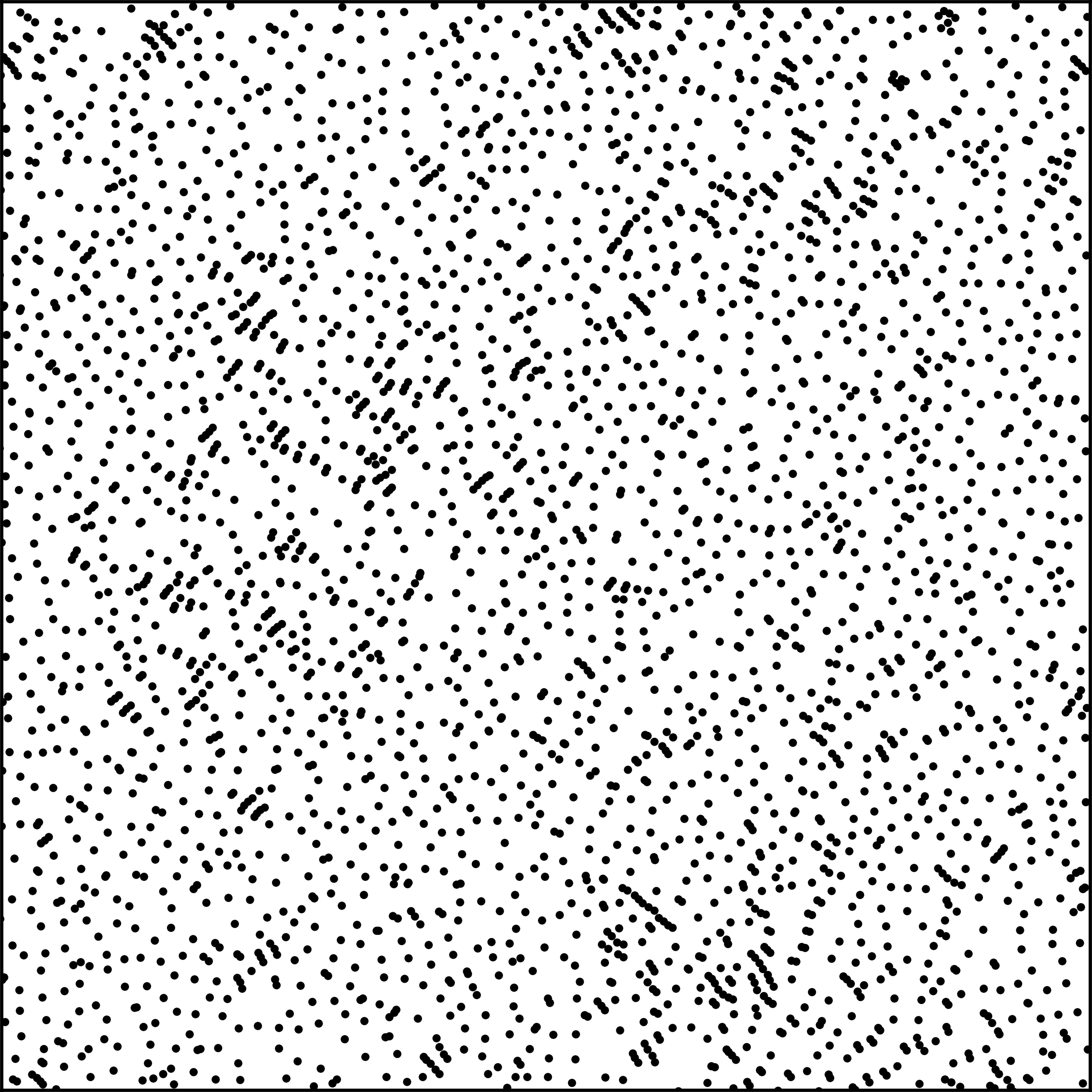} 
    &\includegraphics*[width = 0.24 \textwidth]{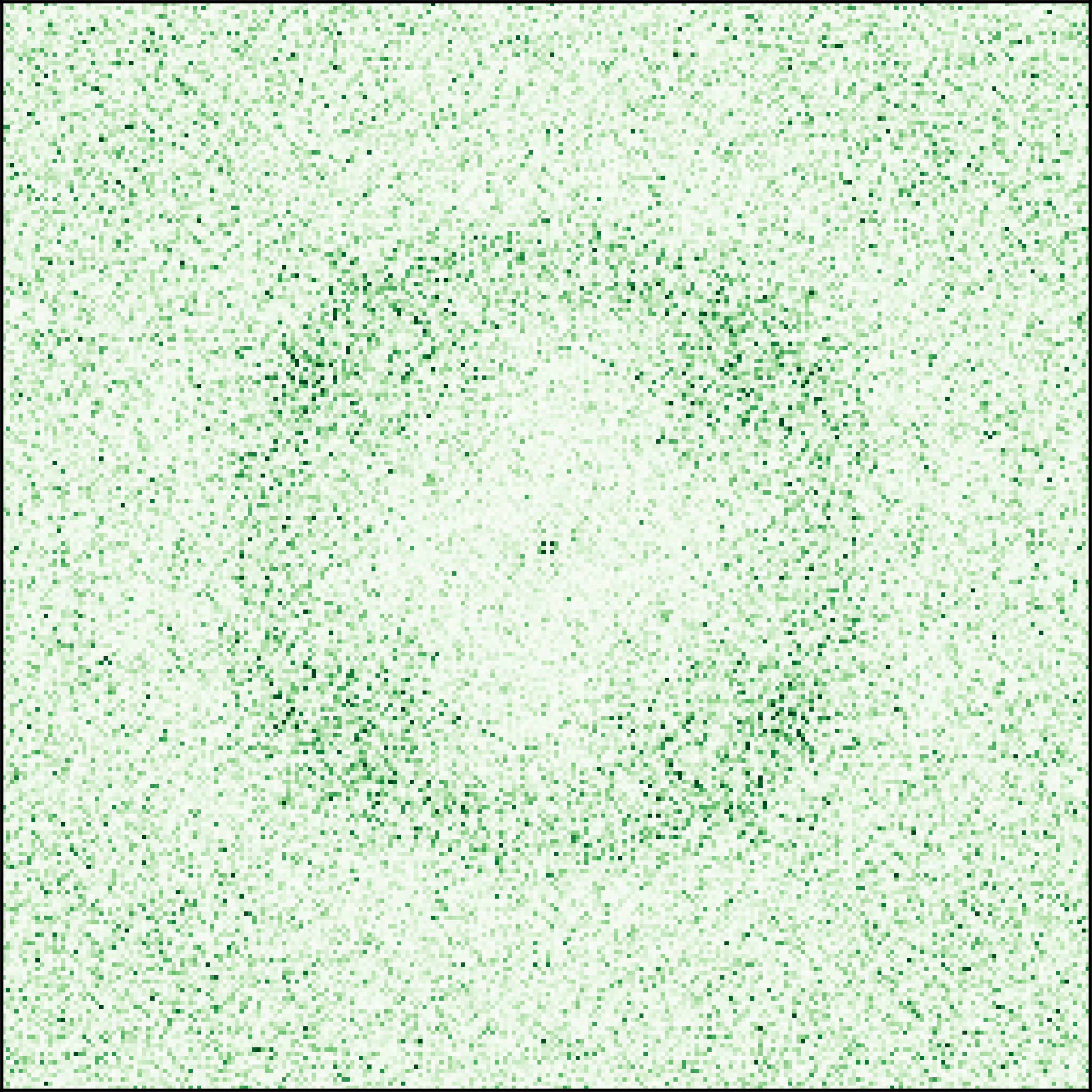} \\
    c) & f)
  \end{tabular}
  \begin{center}
\begin{minipage}{0.95\textwidth}
\caption[short figure description]{a) - c) Point configurations corresponding to extrema of the vorticity field $\ww$ in Figure 1 a) - c), respectively. d) - e) Corresponding spectral density of extrema distribution $\widehat{S}(\mathbf{k})$ to point configurations in a) - c).
}
\label{figSI1}
\end{minipage}
\end{center}
\end{figure*}

Figure \ref{figSI1} a) - c) shows the point configurations of the extracted extrema of the vorticity fields in Figure 1 a) - c). Nearby extrema have been merged. Figure \ref{figSI1} d) - f) show the corresponding spectral density of extrema distribution. 

\begin{figure*}[htb]
  \noindent
  \begin{tabular}{ll}
  \begin{picture}(0.45 \textwidth,0.35 \textwidth)
    \put(0,0){\includegraphics*[width = 0.45 \textwidth ]{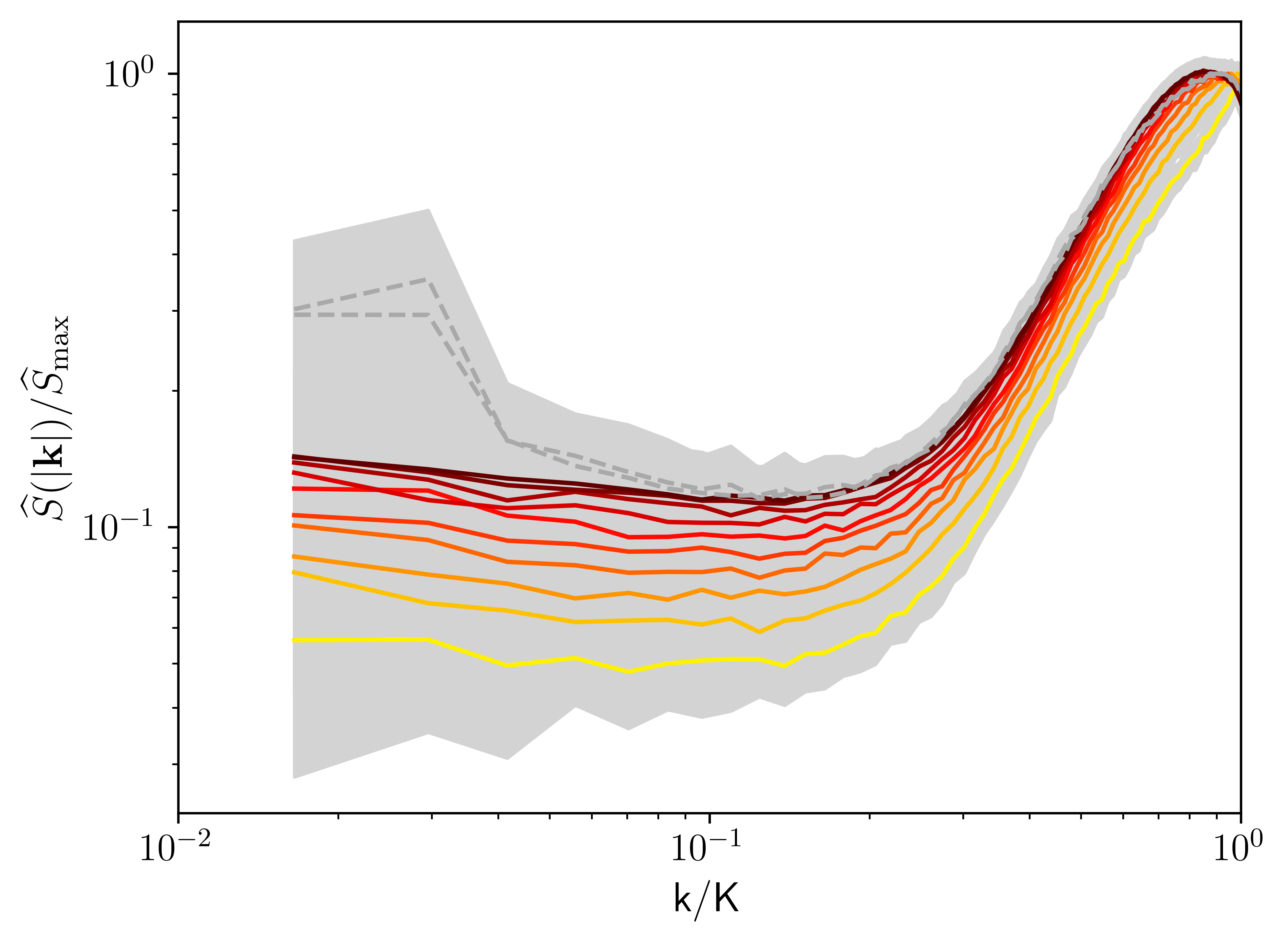}}
    \put(4,8){a)}
     \end{picture}& 
     \begin{picture}(0.45 \textwidth,0.35 \textwidth)
     \put(0,0){\includegraphics*[width = 0.45 \textwidth ]{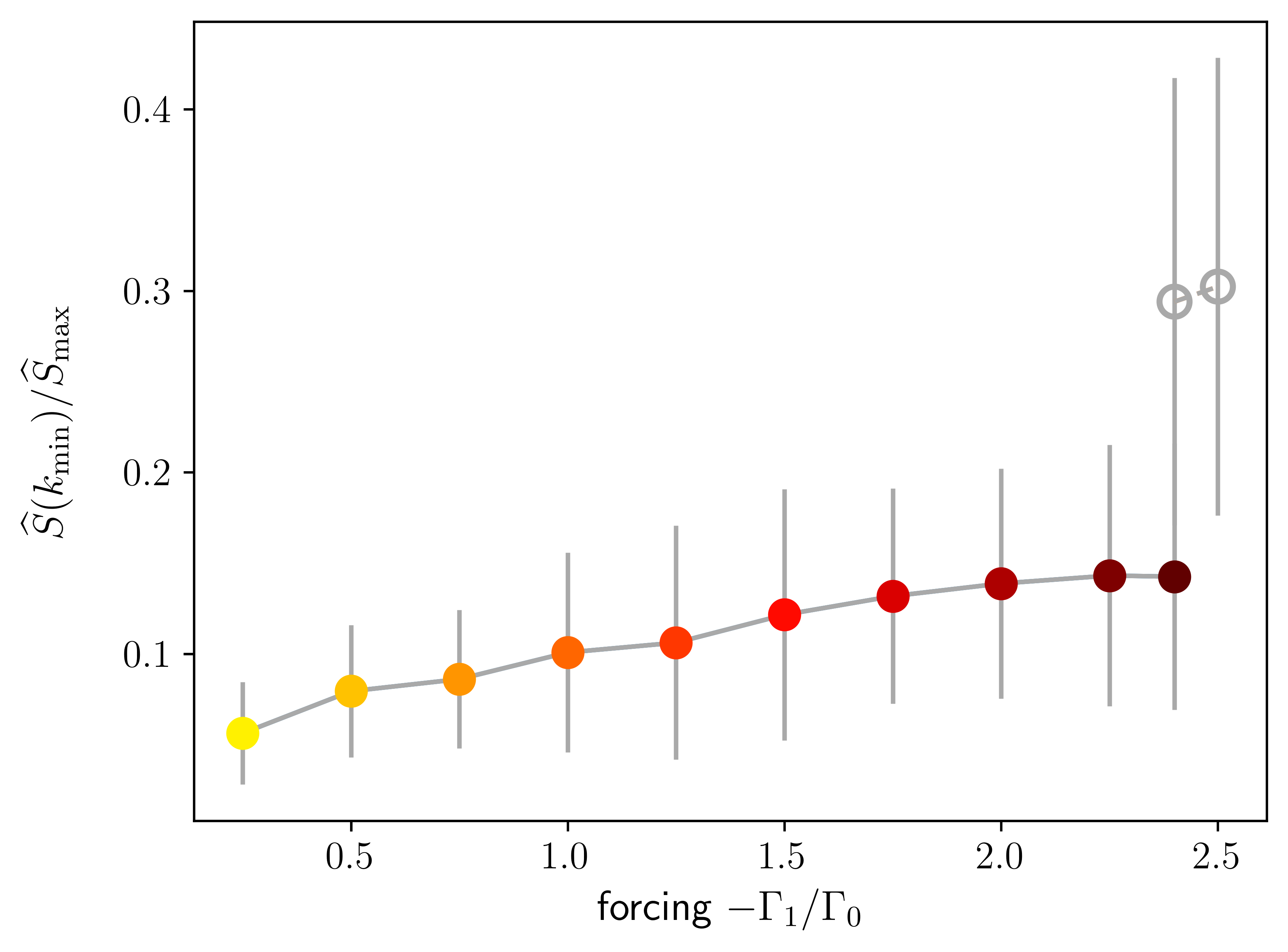}}
     \put(4,8){b)}
     \end{picture}
    %\\
    %a) & b)  
   \end{tabular}
  \begin{center}
\begin{minipage}{0.95\textwidth}
  \caption[short figure description]{
    a) Structure factor $\widehat{S}(\nkk)$ of extracted extrema in vorticity field as a function of $k = \mathbf{k}$ for different forcing $|\Gamma_2 / \Gamma_0|$. Dashed lines correspond to the condensate state. The color coding makes the forcing. The gray shading indicates the standard deviation resulting from the averaging of all time instances. The quantities are rescaled by $K = 36.5$, the center of the region of energy input, and the maximum of $\widehat{S}(\nkk)$. b) $\widehat{S}(\nkk)$ at smallest $\kk$, $\nkk= k_{\rm min}$, normalized by $\widehat{S}_{\rm max}$.
\label{figSI2}
}
\end{minipage}
\end{center}
\end{figure*}

The circular region around the origin in which there is almost no scattering for low activities (Figure \ref{figSI1} d) and e)) provides a first signature for long-range correlations. Slight differences can be seen if compared with Figure \ref{figSI1} f), corresponding to a larger activity above the critical value. More quantitatively, Figure \ref{figSI2} a) and b) show the structure factor $\widehat{S}(|\mathbf{k}|)$ and the value $\widehat{S}( k_{\rm min})$ as a function of forcing strength. While a qualitative change at the critical value is visible, these data do not show clear characteristics of hyperuniformity as the data is too noisy for small $\nkk$. 

\vspace{-10pt}
\section{Numerical validation}

We demonstrate the independence of the simulation results for three selected activities. 
\begin{figure}[htb]
  \noindent
  \begin{center}
\includegraphics*[width = 0.45 \textwidth ]{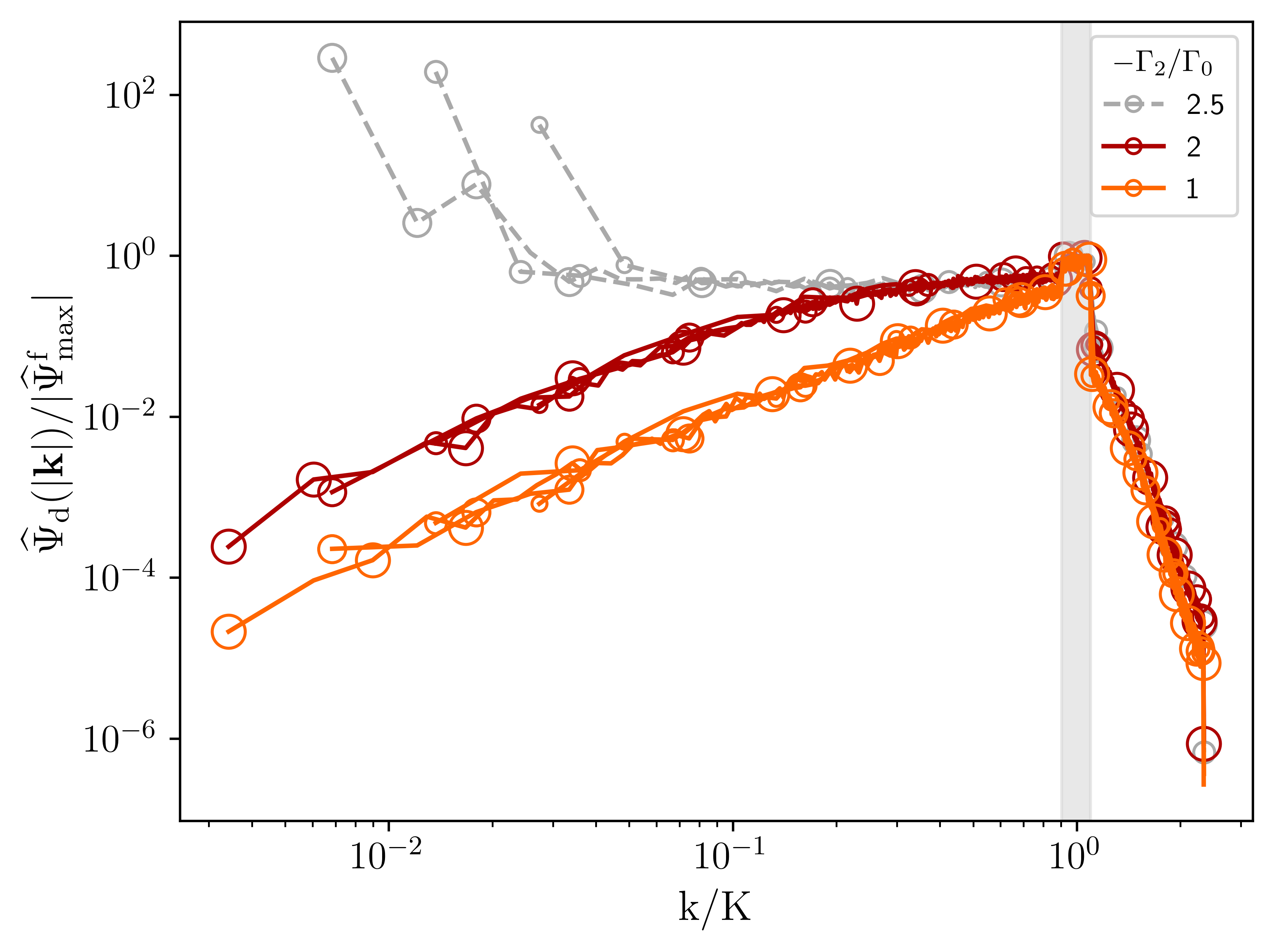}
\caption[short figure description]{
Radial spectra of autocovariance function $\widehat{\Psi}_{\rm d}(|\mathbf{k}|)$ as a function of $k = |\mathbf{k}|$ for different forcing $|\Gamma_2/\Gamma_0|$ and domain sizes. Dashed lines correspond to the condensate state. The gray region marks the energy input. The quantities are rescaled by $K=36.5$, the center of the region of energy input and $\widehat{\Psi}^f_{\rm max}$, the maximum of $\widehat{\Psi}_{\rm d}(|\mathbf{k}|)$ in this region. The size of symbols corresponds to different domain sizes $[-n\pi,n\pi] \times [-n\pi,n\pi]$, with $n=1,2,4,8$ and $n=1$ corresponding to the simulations in the main text.  For $|\Gamma_2/\Gamma_0|=2.5$ only n = 1,2,4 is considered.
%{\color{red} ANMERKUNG: Jeweils 4 Rechnungen, unterschiedliche Größe der Symbole aber gleiche Farbe, zugeordnet zu $\Gamma_2/\Gamma_0$ und alle in einen Plot.}
\label{figSI3}
}
\end{center}
\end{figure}
Figure \ref{figSI3} shows the radial spectra of autocovariance function $\widehat{\Psi}_d(|\mathbf{k}|) / |\widehat{\Psi}_{\rm max}|$ for increased domain sizes of $[-2\pi,2\pi] \times [-2\pi,2\pi]$, $[-4\pi,4\pi] \times [-4\pi,4\pi]$ and $[-8\pi,8\pi] \times [-8\pi,8\pi]$ which are discretized by $2^{9}$, $2^{10}$ and $2^{11}$ points in each direction, respectively. The spectra of the condensate phase, corresponding to $|\Gamma_2/\Gamma_0|=2.5$, are size-dependent as the macroscopic vortex is bound by the domain. However, other properties are not affected. The scaling behavior is the same as in the main text for $n = 1$, and the hyperuniformity metric $H$ further decreases with increasing $n$, confirming nearly hyperuniformity or even effective hyperuniformity and, therefore, type I hyperuniformity for forcing significantly below the critical value. 

\section{Movies}

We provide movies corresponding to the flow field snapshots shown in Figure 1 a) - c). They consider LIC visualization with color coding corresponding to vorticity and show the statistically steady states. The file names correspond to the magnitude of forcing $|\Gamma_2/\Gamma_0|$ {\bf one.mov}, {\bf two.mov} and {\bf twopointfive.mov}. \\

\noindent {\bf Acknowledgements.} MS and AV acknowledge support from the German Research Foundation (DFG) within project FOR3013 "Vector- and tensor-valued surface PDEs". We further acknowledge computing resources provided by ZIH at
TU Dresden within project WIR, and by JSC at FZ J\"ulich, within project PFAMDIS.

%\showacknow{} % Display the acknowledgments section

% Bibliography
\bibliography{ActiveFlowLit}

\begin{thebibliography}{72}%
\makeatletter
\providecommand \@ifxundefined [1]{%
 \@ifx{#1\undefined}
}%
\providecommand \@ifnum [1]{%
 \ifnum #1\expandafter \@firstoftwo
 \else \expandafter \@secondoftwo
 \fi
}%
\providecommand \@ifx [1]{%
 \ifx #1\expandafter \@firstoftwo
 \else \expandafter \@secondoftwo
 \fi
}%
\providecommand \natexlab [1]{#1}%
\providecommand \enquote  [1]{``#1''}%
\providecommand \bibnamefont  [1]{#1}%
\providecommand \bibfnamefont [1]{#1}%
\providecommand \citenamefont [1]{#1}%
\providecommand \href@noop [0]{\@secondoftwo}%
\providecommand \href [0]{\begingroup \@sanitize@url \@href}%
\providecommand \@href[1]{\@@startlink{#1}\@@href}%
\providecommand \@@href[1]{\endgroup#1\@@endlink}%
\providecommand \@sanitize@url [0]{\catcode `\\12\catcode `\$12\catcode
  `\&12\catcode `\#12\catcode `\^12\catcode `\_12\catcode `\%12\relax}%
\providecommand \@@startlink[1]{}%
\providecommand \@@endlink[0]{}%
\providecommand \url  [0]{\begingroup\@sanitize@url \@url }%
\providecommand \@url [1]{\endgroup\@href {#1}{\urlprefix }}%
\providecommand \urlprefix  [0]{URL }%
\providecommand \Eprint [0]{\href }%
\providecommand \doibase [0]{https://doi.org/}%
\providecommand \selectlanguage [0]{\@gobble}%
\providecommand \bibinfo  [0]{\@secondoftwo}%
\providecommand \bibfield  [0]{\@secondoftwo}%
\providecommand \translation [1]{[#1]}%
\providecommand \BibitemOpen [0]{}%
\providecommand \bibitemStop [0]{}%
\providecommand \bibitemNoStop [0]{.\EOS\space}%
\providecommand \EOS [0]{\spacefactor3000\relax}%
\providecommand \BibitemShut  [1]{\csname bibitem#1\endcsname}%
\let\auto@bib@innerbib\@empty
%</preamble>
\bibitem [{\citenamefont {Torquato}\ and\ \citenamefont
  {Stillinger}(2003)}]{TS03}%
  \BibitemOpen
  \bibfield  {author} {\bibinfo {author} {\bibfnamefont {S.}~\bibnamefont
  {Torquato}}\ and\ \bibinfo {author} {\bibfnamefont {F.~H.}\ \bibnamefont
  {Stillinger}},\ }\bibfield  {title} {\bibinfo {title} {{Local density
  fluctuations, hyperuniformity, and order metrics}},\ }\href
  {https://doi.org/10.1103/PhysRevE.68.041113} {\bibfield  {journal} {\bibinfo
  {journal} {Phys. Rev. E}\ }\textbf {\bibinfo {volume} {68}},\ \bibinfo
  {pages} {041113} (\bibinfo {year} {2003})}\BibitemShut {NoStop}%
\bibitem [{\citenamefont {Torquato}(2018)}]{Tor18}%
  \BibitemOpen
  \bibfield  {author} {\bibinfo {author} {\bibfnamefont {S.}~\bibnamefont
  {Torquato}},\ }\bibfield  {title} {\bibinfo {title} {{Hyperuniform states of
  matter}},\ }\href {https://doi.org/10.1016/j.physrep.2018.03.001} {\bibfield
  {journal} {\bibinfo  {journal} {Phys. Reports}\ }\textbf {\bibinfo {volume}
  {745}},\ \bibinfo {pages} {1} (\bibinfo {year} {2018})}\BibitemShut {NoStop}%
\bibitem [{\citenamefont {Wilken}\ \emph {et~al.}(2021)\citenamefont {Wilken},
  \citenamefont {Guerra}, \citenamefont {Levine},\ and\ \citenamefont
  {Chaikin}}]{Wilkenetal_PRL_2021}%
  \BibitemOpen
  \bibfield  {author} {\bibinfo {author} {\bibfnamefont {S.}~\bibnamefont
  {Wilken}}, \bibinfo {author} {\bibfnamefont {R.~E.}\ \bibnamefont {Guerra}},
  \bibinfo {author} {\bibfnamefont {D.}~\bibnamefont {Levine}},\ and\ \bibinfo
  {author} {\bibfnamefont {P.~M.}\ \bibnamefont {Chaikin}},\ }\bibfield
  {title} {\bibinfo {title} {Random close packing as a dynamical phase
  transition},\ }\href {https://doi.org/10.1103/PhysRevLett.127.038002}
  {\bibfield  {journal} {\bibinfo  {journal} {Phys. Rev. Lett.}\ }\textbf
  {\bibinfo {volume} {127}},\ \bibinfo {pages} {038002} (\bibinfo {year}
  {2021})}\BibitemShut {NoStop}%
\bibitem [{\citenamefont {S\'anchez}\ \emph {et~al.}(2023)\citenamefont
  {S\'anchez}, \citenamefont {Maldonado}, \citenamefont {Amig\'o},
  \citenamefont {Nieva}, \citenamefont {Kolton},\ and\ \citenamefont
  {Fasano}}]{SanchezPRB2023}%
  \BibitemOpen
  \bibfield  {author} {\bibinfo {author} {\bibfnamefont {J.~A.}\ \bibnamefont
  {S\'anchez}}, \bibinfo {author} {\bibfnamefont {R.~C.}\ \bibnamefont
  {Maldonado}}, \bibinfo {author} {\bibfnamefont {M.~L.}\ \bibnamefont
  {Amig\'o}}, \bibinfo {author} {\bibfnamefont {G.}~\bibnamefont {Nieva}},
  \bibinfo {author} {\bibfnamefont {A.}~\bibnamefont {Kolton}},\ and\ \bibinfo
  {author} {\bibfnamefont {Y.}~\bibnamefont {Fasano}},\ }\bibfield  {title}
  {\bibinfo {title} {Disordered hyperuniform vortex matter with rhombic
  distortions in fese at low fields},\ }\href
  {https://doi.org/10.1103/PhysRevB.107.094508} {\bibfield  {journal} {\bibinfo
   {journal} {Phys. Rev. B}\ }\textbf {\bibinfo {volume} {107}},\ \bibinfo
  {pages} {094508} (\bibinfo {year} {2023})}\BibitemShut {NoStop}%
\bibitem [{\citenamefont {Zheng}\ \emph {et~al.}(2020)\citenamefont {Zheng},
  \citenamefont {Liu}, \citenamefont {Nan}, \citenamefont {Shen}, \citenamefont
  {Zhang}, \citenamefont {Chen}, \citenamefont {He}, \citenamefont {Xu},
  \citenamefont {Chen}, \citenamefont {Jiao},\ and\ \citenamefont
  {Zhuang}}]{Zhengetal_SA_2020}%
  \BibitemOpen
  \bibfield  {author} {\bibinfo {author} {\bibfnamefont {Y.}~\bibnamefont
  {Zheng}}, \bibinfo {author} {\bibfnamefont {L.}~\bibnamefont {Liu}}, \bibinfo
  {author} {\bibfnamefont {H.}~\bibnamefont {Nan}}, \bibinfo {author}
  {\bibfnamefont {Z.-X.}\ \bibnamefont {Shen}}, \bibinfo {author}
  {\bibfnamefont {G.}~\bibnamefont {Zhang}}, \bibinfo {author} {\bibfnamefont
  {D.}~\bibnamefont {Chen}}, \bibinfo {author} {\bibfnamefont {L.}~\bibnamefont
  {He}}, \bibinfo {author} {\bibfnamefont {W.}~\bibnamefont {Xu}}, \bibinfo
  {author} {\bibfnamefont {M.}~\bibnamefont {Chen}}, \bibinfo {author}
  {\bibfnamefont {Y.}~\bibnamefont {Jiao}},\ and\ \bibinfo {author}
  {\bibfnamefont {H.}~\bibnamefont {Zhuang}},\ }\bibfield  {title} {\bibinfo
  {title} {Disordered hyperuniformity in two-dimensional amorphous silica},\
  }\href {https://doi.org/10.1126/sciadv.aba0826} {\bibfield  {journal}
  {\bibinfo  {journal} {Sci. Adv.}\ }\textbf {\bibinfo {volume} {6}},\ \bibinfo
  {pages} {eaba0826} (\bibinfo {year} {2020})}\BibitemShut {NoStop}%
\bibitem [{\citenamefont {Chen}\ \emph {et~al.}(2021)\citenamefont {Chen},
  \citenamefont {Zheng}, \citenamefont {Liu}, \citenamefont {Zhang},
  \citenamefont {Chen}, \citenamefont {Jiao},\ and\ \citenamefont
  {Zhuang}}]{Chenetal_PNAS_2021}%
  \BibitemOpen
  \bibfield  {author} {\bibinfo {author} {\bibfnamefont {D.}~\bibnamefont
  {Chen}}, \bibinfo {author} {\bibfnamefont {Y.}~\bibnamefont {Zheng}},
  \bibinfo {author} {\bibfnamefont {L.}~\bibnamefont {Liu}}, \bibinfo {author}
  {\bibfnamefont {G.}~\bibnamefont {Zhang}}, \bibinfo {author} {\bibfnamefont
  {M.}~\bibnamefont {Chen}}, \bibinfo {author} {\bibfnamefont {Y.}~\bibnamefont
  {Jiao}},\ and\ \bibinfo {author} {\bibfnamefont {H.}~\bibnamefont {Zhuang}},\
  }\bibfield  {title} {\bibinfo {title} {{Stone-Wales} defects preserve
  hyperuniformity in amorphous two-dimensional networks},\ }\href
  {https://doi.org/10.1073/pnas.2016862118} {\bibfield  {journal} {\bibinfo
  {journal} {Proc. Nat. Acad. Sci. (USA)}\ }\textbf {\bibinfo {volume} {118}},\
  \bibinfo {pages} {e2016862118} (\bibinfo {year} {2021})}\BibitemShut
  {NoStop}%
\bibitem [{\citenamefont {Ghosh}\ and\ \citenamefont
  {Lebowitz}(2017)}]{Ghoshetal_JSP_2017}%
  \BibitemOpen
  \bibfield  {author} {\bibinfo {author} {\bibfnamefont {S.}~\bibnamefont
  {Ghosh}}\ and\ \bibinfo {author} {\bibfnamefont {J.}~\bibnamefont
  {Lebowitz}},\ }\bibfield  {title} {\bibinfo {title} {Number rigidity in
  superhomogeneous random point fields},\ }\href
  {https://doi.org/10.1007/s10955-016-1633-6} {\bibfield  {journal} {\bibinfo
  {journal} {J. Stat. Phys.}\ }\textbf {\bibinfo {volume} {166}},\ \bibinfo
  {pages} {1016} (\bibinfo {year} {2017})}\BibitemShut {NoStop}%
\bibitem [{\citenamefont {Chatterjee}(2019)}]{Chatterjee_PTRF_2019}%
  \BibitemOpen
  \bibfield  {author} {\bibinfo {author} {\bibfnamefont {S.}~\bibnamefont
  {Chatterjee}},\ }\bibfield  {title} {\bibinfo {title} {Rigidity of the
  three-dimensional hierarchical coulomb gas},\ }\href
  {https://doi.org/10.1007/s00440-019-00912-6} {\bibfield  {journal} {\bibinfo
  {journal} {Prob. Theo. Rel. Fields}\ }\textbf {\bibinfo {volume} {175}},\
  \bibinfo {pages} {1123} (\bibinfo {year} {2019})}\BibitemShut {NoStop}%
\bibitem [{\citenamefont {Florescu}\ \emph {et~al.}(2009)\citenamefont
  {Florescu}, \citenamefont {Torquato},\ and\ \citenamefont
  {Steinhardt}}]{Fetal2009}%
  \BibitemOpen
  \bibfield  {author} {\bibinfo {author} {\bibfnamefont {M.}~\bibnamefont
  {Florescu}}, \bibinfo {author} {\bibfnamefont {S.}~\bibnamefont {Torquato}},\
  and\ \bibinfo {author} {\bibfnamefont {P.~J.}\ \bibnamefont {Steinhardt}},\
  }\bibfield  {title} {\bibinfo {title} {Designer disordered materials with
  large, complete photonic band gaps},\ }\href
  {https://doi.org/10.1073/pnas.0907744106} {\bibfield  {journal} {\bibinfo
  {journal} {Proc. Nat. Acad. Sci. (USA)}\ }\textbf {\bibinfo {volume} {106}},\
  \bibinfo {pages} {20658} (\bibinfo {year} {2009})}\BibitemShut {NoStop}%
\bibitem [{\citenamefont {Froufe-P{\'e}rez}\ \emph {et~al.}(2017)\citenamefont
  {Froufe-P{\'e}rez}, \citenamefont {Engel}, \citenamefont {S{\'a}enz},\ and\
  \citenamefont {Scheffold}}]{froufe2017band}%
  \BibitemOpen
  \bibfield  {author} {\bibinfo {author} {\bibfnamefont {L.~S.}\ \bibnamefont
  {Froufe-P{\'e}rez}}, \bibinfo {author} {\bibfnamefont {M.}~\bibnamefont
  {Engel}}, \bibinfo {author} {\bibfnamefont {J.~J.}\ \bibnamefont
  {S{\'a}enz}},\ and\ \bibinfo {author} {\bibfnamefont {F.}~\bibnamefont
  {Scheffold}},\ }\bibfield  {title} {\bibinfo {title} {Band gap formation and
  anderson localization in disordered photonic materials with structural
  correlations},\ }\href {https://doi.org/10.1073/pnas.1705130114} {\bibfield
  {journal} {\bibinfo  {journal} {Proc. Nat. Acad. Sci. (USA)}\ }\textbf
  {\bibinfo {volume} {114}},\ \bibinfo {pages} {9570} (\bibinfo {year}
  {2017})}\BibitemShut {NoStop}%
\bibitem [{\citenamefont {Mitchell}\ \emph {et~al.}(2018)\citenamefont
  {Mitchell}, \citenamefont {Nash}, \citenamefont {Hexner}, \citenamefont
  {Turner},\ and\ \citenamefont {Irvine}}]{mitchell2018amorphous}%
  \BibitemOpen
  \bibfield  {author} {\bibinfo {author} {\bibfnamefont {N.~P.}\ \bibnamefont
  {Mitchell}}, \bibinfo {author} {\bibfnamefont {L.~M.}\ \bibnamefont {Nash}},
  \bibinfo {author} {\bibfnamefont {D.}~\bibnamefont {Hexner}}, \bibinfo
  {author} {\bibfnamefont {A.~M.}\ \bibnamefont {Turner}},\ and\ \bibinfo
  {author} {\bibfnamefont {W.~T.}\ \bibnamefont {Irvine}},\ }\bibfield  {title}
  {\bibinfo {title} {Amorphous topological insulators constructed from random
  point sets},\ }\href {https://doi.org/10.1038/s41567-017-0024-5} {\bibfield
  {journal} {\bibinfo  {journal} {Nature Physics}\ }\textbf {\bibinfo {volume}
  {14}},\ \bibinfo {pages} {380} (\bibinfo {year} {2018})}\BibitemShut
  {NoStop}%
\bibitem [{\citenamefont {Gerasimenko}\ \emph {et~al.}(2019)\citenamefont
  {Gerasimenko}, \citenamefont {Vaskivskyi}, \citenamefont {Litskevich},
  \citenamefont {Ravnik}, \citenamefont {Vodeb}, \citenamefont {Diego},
  \citenamefont {Kabanov},\ and\ \citenamefont
  {Mihailovic}}]{gerasimenko2019quantum}%
  \BibitemOpen
  \bibfield  {author} {\bibinfo {author} {\bibfnamefont {Y.~A.}\ \bibnamefont
  {Gerasimenko}}, \bibinfo {author} {\bibfnamefont {I.}~\bibnamefont
  {Vaskivskyi}}, \bibinfo {author} {\bibfnamefont {M.}~\bibnamefont
  {Litskevich}}, \bibinfo {author} {\bibfnamefont {J.}~\bibnamefont {Ravnik}},
  \bibinfo {author} {\bibfnamefont {J.}~\bibnamefont {Vodeb}}, \bibinfo
  {author} {\bibfnamefont {M.}~\bibnamefont {Diego}}, \bibinfo {author}
  {\bibfnamefont {V.}~\bibnamefont {Kabanov}},\ and\ \bibinfo {author}
  {\bibfnamefont {D.}~\bibnamefont {Mihailovic}},\ }\bibfield  {title}
  {\bibinfo {title} {Quantum jamming transition to a correlated electron glass
  in 1t-tas2},\ }\href {https://doi.org/10.1038/s41563-019-0423-3} {\bibfield
  {journal} {\bibinfo  {journal} {Nature Materials}\ }\textbf {\bibinfo
  {volume} {18}},\ \bibinfo {pages} {1078} (\bibinfo {year}
  {2019})}\BibitemShut {NoStop}%
\bibitem [{\citenamefont {Yu}\ \emph {et~al.}(2021)\citenamefont {Yu},
  \citenamefont {Qiu}, \citenamefont {Chong}, \citenamefont {Torquato},\ and\
  \citenamefont {Park}}]{yu2021engineered}%
  \BibitemOpen
  \bibfield  {author} {\bibinfo {author} {\bibfnamefont {S.}~\bibnamefont
  {Yu}}, \bibinfo {author} {\bibfnamefont {C.-W.}\ \bibnamefont {Qiu}},
  \bibinfo {author} {\bibfnamefont {Y.}~\bibnamefont {Chong}}, \bibinfo
  {author} {\bibfnamefont {S.}~\bibnamefont {Torquato}},\ and\ \bibinfo
  {author} {\bibfnamefont {N.}~\bibnamefont {Park}},\ }\bibfield  {title}
  {\bibinfo {title} {Engineered disorder in photonics},\ }\href
  {https://doi.org/10.1038/7257780a} {\bibfield  {journal} {\bibinfo  {journal}
  {Nat. Rev. Mater.}\ }\textbf {\bibinfo {volume} {6}},\ \bibinfo {pages} {226}
  (\bibinfo {year} {2021})}\BibitemShut {NoStop}%
\bibitem [{\citenamefont {Vynck}\ \emph {et~al.}(2023)\citenamefont {Vynck},
  \citenamefont {Pierrat}, \citenamefont {Carminati}, \citenamefont
  {Froufe-P\'erez}, \citenamefont {Scheffold}, \citenamefont {Sapienza},
  \citenamefont {Vignolini},\ and\ \citenamefont {S\'aenz}}]{Vynck2023}%
  \BibitemOpen
  \bibfield  {author} {\bibinfo {author} {\bibfnamefont {K.}~\bibnamefont
  {Vynck}}, \bibinfo {author} {\bibfnamefont {R.}~\bibnamefont {Pierrat}},
  \bibinfo {author} {\bibfnamefont {R.}~\bibnamefont {Carminati}}, \bibinfo
  {author} {\bibfnamefont {L.~S.}\ \bibnamefont {Froufe-P\'erez}}, \bibinfo
  {author} {\bibfnamefont {F.}~\bibnamefont {Scheffold}}, \bibinfo {author}
  {\bibfnamefont {R.}~\bibnamefont {Sapienza}}, \bibinfo {author}
  {\bibfnamefont {S.}~\bibnamefont {Vignolini}},\ and\ \bibinfo {author}
  {\bibfnamefont {J.~J.}\ \bibnamefont {S\'aenz}},\ }\bibfield  {title}
  {\bibinfo {title} {Light in correlated disordered media},\ }\href
  {https://doi.org/10.1103/RevModPhys.95.045003} {\bibfield  {journal}
  {\bibinfo  {journal} {Rev. Mod. Phys.}\ }\textbf {\bibinfo {volume} {95}},\
  \bibinfo {pages} {045003} (\bibinfo {year} {2023})}\BibitemShut {NoStop}%
\bibitem [{\citenamefont {Tavakoli}\ \emph {et~al.}(2022)\citenamefont
  {Tavakoli}, \citenamefont {Spalding}, \citenamefont {Lambertz}, \citenamefont
  {Koppejan}, \citenamefont {Gkantzounis}, \citenamefont {Wan}, \citenamefont
  {Röhrich}, \citenamefont {Kontoleta}, \citenamefont {Koenderink},
  \citenamefont {Sapienza}, \citenamefont {Florescu},\ and\ \citenamefont
  {Alarcon-Llado}}]{Tavakoli_ACS_2022}%
  \BibitemOpen
  \bibfield  {author} {\bibinfo {author} {\bibfnamefont {N.}~\bibnamefont
  {Tavakoli}}, \bibinfo {author} {\bibfnamefont {R.}~\bibnamefont {Spalding}},
  \bibinfo {author} {\bibfnamefont {A.}~\bibnamefont {Lambertz}}, \bibinfo
  {author} {\bibfnamefont {P.}~\bibnamefont {Koppejan}}, \bibinfo {author}
  {\bibfnamefont {G.}~\bibnamefont {Gkantzounis}}, \bibinfo {author}
  {\bibfnamefont {C.}~\bibnamefont {Wan}}, \bibinfo {author} {\bibfnamefont
  {R.}~\bibnamefont {Röhrich}}, \bibinfo {author} {\bibfnamefont
  {E.}~\bibnamefont {Kontoleta}}, \bibinfo {author} {\bibfnamefont {A.~F.}\
  \bibnamefont {Koenderink}}, \bibinfo {author} {\bibfnamefont
  {R.}~\bibnamefont {Sapienza}}, \bibinfo {author} {\bibfnamefont
  {M.}~\bibnamefont {Florescu}},\ and\ \bibinfo {author} {\bibfnamefont
  {E.}~\bibnamefont {Alarcon-Llado}},\ }\bibfield  {title} {\bibinfo {title}
  {Over 65\% sunlight absorption in a 1 $\mu m$ si slab with hyperuniform
  texture},\ }\href {https://doi.org/10.1021/acsphotonics.1c01668} {\bibfield
  {journal} {\bibinfo  {journal} {ACS Photonics}\ }\textbf {\bibinfo {volume}
  {9}},\ \bibinfo {pages} {1206} (\bibinfo {year} {2022})}\BibitemShut
  {NoStop}%
\bibitem [{\citenamefont {Jiao}\ \emph {et~al.}(2014)\citenamefont {Jiao},
  \citenamefont {Lau}, \citenamefont {Hatzikirou}, \citenamefont
  {Meyer-Hermann}, \citenamefont {Corbo},\ and\ \citenamefont
  {Torquato}}]{Jiaoetal_PRE_2014}%
  \BibitemOpen
  \bibfield  {author} {\bibinfo {author} {\bibfnamefont {Y.}~\bibnamefont
  {Jiao}}, \bibinfo {author} {\bibfnamefont {T.}~\bibnamefont {Lau}}, \bibinfo
  {author} {\bibfnamefont {H.}~\bibnamefont {Hatzikirou}}, \bibinfo {author}
  {\bibfnamefont {M.}~\bibnamefont {Meyer-Hermann}}, \bibinfo {author}
  {\bibfnamefont {J.~C.}\ \bibnamefont {Corbo}},\ and\ \bibinfo {author}
  {\bibfnamefont {S.}~\bibnamefont {Torquato}},\ }\bibfield  {title} {\bibinfo
  {title} {Avian photoreceptor patterns represent a disordered hyperuniform
  solution to a multiscale packing problem},\ }\href
  {https://doi.org/10.1103/PhysRevE.89.022721} {\bibfield  {journal} {\bibinfo
  {journal} {Phys. Rev. E}\ }\textbf {\bibinfo {volume} {89}},\ \bibinfo
  {pages} {022721} (\bibinfo {year} {2014})}\BibitemShut {NoStop}%
\bibitem [{\citenamefont {Mayer}\ \emph {et~al.}(2015)\citenamefont {Mayer},
  \citenamefont {Balasubramanian}, \citenamefont {Mora},\ and\ \citenamefont
  {Walczak}}]{Mayeretal_PNAS_2015}%
  \BibitemOpen
  \bibfield  {author} {\bibinfo {author} {\bibfnamefont {A.}~\bibnamefont
  {Mayer}}, \bibinfo {author} {\bibfnamefont {V.}~\bibnamefont
  {Balasubramanian}}, \bibinfo {author} {\bibfnamefont {T.}~\bibnamefont
  {Mora}},\ and\ \bibinfo {author} {\bibfnamefont {A.~M.}\ \bibnamefont
  {Walczak}},\ }\bibfield  {title} {\bibinfo {title} {How a well-adapted immune
  system is organized},\ }\href {https://doi.org/10.1073/pnas.1421827112}
  {\bibfield  {journal} {\bibinfo  {journal} {Proc. Nat. Acad. Sci. (USA)}\
  }\textbf {\bibinfo {volume} {112}},\ \bibinfo {pages} {5950} (\bibinfo {year}
  {2015})}\BibitemShut {NoStop}%
\bibitem [{\citenamefont {Ramaswamy}(2010)}]{RamaswanyARCMP2010}%
  \BibitemOpen
  \bibfield  {author} {\bibinfo {author} {\bibfnamefont {S.}~\bibnamefont
  {Ramaswamy}},\ }\bibfield  {title} {\bibinfo {title} {The mechanics and
  statistics of active matter},\ }\href
  {https://doi.org/10.1146/annurev-conmatphys-070909-104101} {\bibfield
  {journal} {\bibinfo  {journal} {Annual Review of Condensed Matter Physics}\
  }\textbf {\bibinfo {volume} {1}},\ \bibinfo {pages} {323} (\bibinfo {year}
  {2010})}\BibitemShut {NoStop}%
\bibitem [{\citenamefont {Marchetti}\ \emph {et~al.}(2013)\citenamefont
  {Marchetti}, \citenamefont {Joanny}, \citenamefont {Ramaswamy}, \citenamefont
  {Liverpool}, \citenamefont {Prost}, \citenamefont {Rao},\ and\ \citenamefont
  {Simha}}]{Marchetti2013}%
  \BibitemOpen
  \bibfield  {author} {\bibinfo {author} {\bibfnamefont {M.~C.}\ \bibnamefont
  {Marchetti}}, \bibinfo {author} {\bibfnamefont {J.~F.}\ \bibnamefont
  {Joanny}}, \bibinfo {author} {\bibfnamefont {S.}~\bibnamefont {Ramaswamy}},
  \bibinfo {author} {\bibfnamefont {T.~B.}\ \bibnamefont {Liverpool}}, \bibinfo
  {author} {\bibfnamefont {J.}~\bibnamefont {Prost}}, \bibinfo {author}
  {\bibfnamefont {M.}~\bibnamefont {Rao}},\ and\ \bibinfo {author}
  {\bibfnamefont {R.~A.}\ \bibnamefont {Simha}},\ }\bibfield  {title} {\bibinfo
  {title} {Hydrodynamics of soft active matter},\ }\href
  {https://doi.org/10.1103/RevModPhys.85.1143} {\bibfield  {journal} {\bibinfo
  {journal} {Reviews of Modern Physics}\ }\textbf {\bibinfo {volume} {85}},\
  \bibinfo {pages} {1143} (\bibinfo {year} {2013})}\BibitemShut {NoStop}%
\bibitem [{\citenamefont {Ramaswamy}\ \emph {et~al.}(2003)\citenamefont
  {Ramaswamy}, \citenamefont {Simha},\ and\ \citenamefont
  {Toner}}]{Ramaswamyetal_EPL_2003}%
  \BibitemOpen
  \bibfield  {author} {\bibinfo {author} {\bibfnamefont {S.}~\bibnamefont
  {Ramaswamy}}, \bibinfo {author} {\bibfnamefont {R.~A.}\ \bibnamefont
  {Simha}},\ and\ \bibinfo {author} {\bibfnamefont {J.}~\bibnamefont {Toner}},\
  }\bibfield  {title} {\bibinfo {title} {Active nematics on a substrate: Giant
  number fluctuations and long-time tails},\ }\href
  {https://doi.org/10.1209/epl/i2003-00346-7} {\bibfield  {journal} {\bibinfo
  {journal} {EPL}\ }\textbf {\bibinfo {volume} {62}},\ \bibinfo {pages} {196}
  (\bibinfo {year} {2003})}\BibitemShut {NoStop}%
\bibitem [{\citenamefont {Palacci}\ \emph {et~al.}(2013)\citenamefont
  {Palacci}, \citenamefont {Sacanna}, \citenamefont {Steinberg}, \citenamefont
  {Pine},\ and\ \citenamefont {Chaikin}}]{Palacci2013}%
  \BibitemOpen
  \bibfield  {author} {\bibinfo {author} {\bibfnamefont {J.}~\bibnamefont
  {Palacci}}, \bibinfo {author} {\bibfnamefont {S.}~\bibnamefont {Sacanna}},
  \bibinfo {author} {\bibfnamefont {A.~P.}\ \bibnamefont {Steinberg}}, \bibinfo
  {author} {\bibfnamefont {D.~J.}\ \bibnamefont {Pine}},\ and\ \bibinfo
  {author} {\bibfnamefont {P.~M.}\ \bibnamefont {Chaikin}},\ }\bibfield
  {title} {\bibinfo {title} {Living crystals of light-activated colloidal
  surfers},\ }\href {https://doi.org/10.1126/science.1230020} {\bibfield
  {journal} {\bibinfo  {journal} {Science}\ }\textbf {\bibinfo {volume}
  {339}},\ \bibinfo {pages} {936} (\bibinfo {year} {2013})}\BibitemShut
  {NoStop}%
\bibitem [{\citenamefont {Alaimo}\ \emph {et~al.}(2016)\citenamefont {Alaimo},
  \citenamefont {Praetorius},\ and\ \citenamefont
  {Voigt}}]{Alaimoetal_NJP_2016}%
  \BibitemOpen
  \bibfield  {author} {\bibinfo {author} {\bibfnamefont {F.}~\bibnamefont
  {Alaimo}}, \bibinfo {author} {\bibfnamefont {S.}~\bibnamefont {Praetorius}},\
  and\ \bibinfo {author} {\bibfnamefont {A.}~\bibnamefont {Voigt}},\ }\bibfield
   {title} {\bibinfo {title} {A microscopic field theoretical approach for
  active systems},\ }\href {https://doi.org/10.1088/1367-2630/18/8/083008}
  {\bibfield  {journal} {\bibinfo  {journal} {New J. Phys.}\ }\textbf {\bibinfo
  {volume} {18}},\ \bibinfo {pages} {083008} (\bibinfo {year}
  {2016})}\BibitemShut {NoStop}%
\bibitem [{\citenamefont {Buttinoni}\ \emph {et~al.}(2013)\citenamefont
  {Buttinoni}, \citenamefont {Bialk\'e}, \citenamefont {K\"ummel},
  \citenamefont {L\"owen}, \citenamefont {Bechinger},\ and\ \citenamefont
  {Speck}}]{Buttinonietal_PRL_2013}%
  \BibitemOpen
  \bibfield  {author} {\bibinfo {author} {\bibfnamefont {I.}~\bibnamefont
  {Buttinoni}}, \bibinfo {author} {\bibfnamefont {J.}~\bibnamefont {Bialk\'e}},
  \bibinfo {author} {\bibfnamefont {F.}~\bibnamefont {K\"ummel}}, \bibinfo
  {author} {\bibfnamefont {H.}~\bibnamefont {L\"owen}}, \bibinfo {author}
  {\bibfnamefont {C.}~\bibnamefont {Bechinger}},\ and\ \bibinfo {author}
  {\bibfnamefont {T.}~\bibnamefont {Speck}},\ }\bibfield  {title} {\bibinfo
  {title} {Dynamical clustering and phase separation in suspensions of
  self-propelled colloidal particles},\ }\href@noop {} {\bibfield  {journal}
  {\bibinfo  {journal} {Phys. Rev. Lett.}\ }\textbf {\bibinfo {volume} {110}},\
  \bibinfo {pages} {238301} (\bibinfo {year} {2013})}\BibitemShut {NoStop}%
\bibitem [{\citenamefont {Cates}\ and\ \citenamefont
  {Tailleur}(2015)}]{Catesetal_ARCMP_2015}%
  \BibitemOpen
  \bibfield  {author} {\bibinfo {author} {\bibfnamefont {M.~E.}\ \bibnamefont
  {Cates}}\ and\ \bibinfo {author} {\bibfnamefont {J.}~\bibnamefont
  {Tailleur}},\ }\bibfield  {title} {\bibinfo {title} {Motility-induced phase
  separation},\ }\href
  {https://doi.org/10.1146/annurev-conmatphys-031214-014710} {\bibfield
  {journal} {\bibinfo  {journal} {Ann. Rev. Cond. Matt. Phys.}\ }\textbf
  {\bibinfo {volume} {6}},\ \bibinfo {pages} {219} (\bibinfo {year}
  {2015})}\BibitemShut {NoStop}%
\bibitem [{\citenamefont {Lei}\ and\ \citenamefont
  {Ni}(2019)}]{Leietal_PNAS_2019}%
  \BibitemOpen
  \bibfield  {author} {\bibinfo {author} {\bibfnamefont {Q.-L.}\ \bibnamefont
  {Lei}}\ and\ \bibinfo {author} {\bibfnamefont {R.}~\bibnamefont {Ni}},\
  }\bibfield  {title} {\bibinfo {title} {Hydrodynamics of random-organizing
  hyperuniform fluids},\ }\href {https://doi.org/10.1073/pnas.1911596116}
  {\bibfield  {journal} {\bibinfo  {journal} {Proc. Nat. Acad. Sci. (USA)}\
  }\textbf {\bibinfo {volume} {116}},\ \bibinfo {pages} {22983} (\bibinfo
  {year} {2019})}\BibitemShut {NoStop}%
\bibitem [{\citenamefont {Lei}\ \emph {et~al.}(2019)\citenamefont {Lei},
  \citenamefont {Ciamarra},\ and\ \citenamefont {Ni}}]{Leietal_SA_2019}%
  \BibitemOpen
  \bibfield  {author} {\bibinfo {author} {\bibfnamefont {Q.-L.}\ \bibnamefont
  {Lei}}, \bibinfo {author} {\bibfnamefont {M.~P.}\ \bibnamefont {Ciamarra}},\
  and\ \bibinfo {author} {\bibfnamefont {R.}~\bibnamefont {Ni}},\ }\bibfield
  {title} {\bibinfo {title} {Nonequilibrium strongly hyperuniform fluids of
  circle active particles with large local density fluctuations},\ }\href
  {https://doi.org/10.1126/sciadv.aau7423} {\bibfield  {journal} {\bibinfo
  {journal} {Sci. Adv.}\ }\textbf {\bibinfo {volume} {5}},\ \bibinfo {pages}
  {eaau7423} (\bibinfo {year} {2019})}\BibitemShut {NoStop}%
\bibitem [{\citenamefont {Huang}\ \emph {et~al.}(2021)\citenamefont {Huang},
  \citenamefont {Hu}, \citenamefont {Yang}, \citenamefont {Liu},\ and\
  \citenamefont {Zhang}}]{Huangetal_PNAS_2021}%
  \BibitemOpen
  \bibfield  {author} {\bibinfo {author} {\bibfnamefont {M.}~\bibnamefont
  {Huang}}, \bibinfo {author} {\bibfnamefont {W.}~\bibnamefont {Hu}}, \bibinfo
  {author} {\bibfnamefont {S.}~\bibnamefont {Yang}}, \bibinfo {author}
  {\bibfnamefont {Q.-X.}\ \bibnamefont {Liu}},\ and\ \bibinfo {author}
  {\bibfnamefont {H.~P.}\ \bibnamefont {Zhang}},\ }\bibfield  {title} {\bibinfo
  {title} {Circular swimming motility and disordered hyperuniform state in an
  algae system},\ }\href {https://doi.org/10.1073/pnas.2100493118} {\bibfield
  {journal} {\bibinfo  {journal} {Proc. Nat. Acad. Sci. (USA)}\ }\textbf
  {\bibinfo {volume} {118}},\ \bibinfo {pages} {e2100493118} (\bibinfo {year}
  {2021})}\BibitemShut {NoStop}%
\bibitem [{\citenamefont {Oppenheimer}\ \emph {et~al.}(2022)\citenamefont
  {Oppenheimer}, \citenamefont {Stein}, \citenamefont {Ben~Zion},\ and\
  \citenamefont {Shelley}}]{Oppenheimeretal_NC_2022}%
  \BibitemOpen
  \bibfield  {author} {\bibinfo {author} {\bibfnamefont {N.}~\bibnamefont
  {Oppenheimer}}, \bibinfo {author} {\bibfnamefont {D.~B.}\ \bibnamefont
  {Stein}}, \bibinfo {author} {\bibfnamefont {M.~Y.}\ \bibnamefont
  {Ben~Zion}},\ and\ \bibinfo {author} {\bibfnamefont {M.~J.}\ \bibnamefont
  {Shelley}},\ }\bibfield  {title} {\bibinfo {title} {Hyperuniformity and phase
  enrichment in vortex and rotor assemblies},\ }\href
  {https://doi.org/10.1038/s41467-022-28375-9} {\bibfield  {journal} {\bibinfo
  {journal} {Nature Commun.}\ }\textbf {\bibinfo {volume} {13}},\ \bibinfo
  {pages} {804} (\bibinfo {year} {2022})}\BibitemShut {NoStop}%
\bibitem [{\citenamefont {Slomka}\ and\ \citenamefont
  {Dunkel}(2015)}]{Slomkaetal_EPJSP_2015}%
  \BibitemOpen
  \bibfield  {author} {\bibinfo {author} {\bibfnamefont {J.}~\bibnamefont
  {Slomka}}\ and\ \bibinfo {author} {\bibfnamefont {J.}~\bibnamefont
  {Dunkel}},\ }\bibfield  {title} {\bibinfo {title} {Generalized navier-stokes
  equations for active suspensions},\ }\href
  {https://doi.org/10.1140/epjst/e2015-02463-2} {\bibfield  {journal} {\bibinfo
   {journal} {Europ. Phys. J. ST}\ }\textbf {\bibinfo {volume} {224}},\
  \bibinfo {pages} {1349} (\bibinfo {year} {2015})}\BibitemShut {NoStop}%
\bibitem [{\citenamefont {Linkmann}\ \emph {et~al.}(2019)\citenamefont
  {Linkmann}, \citenamefont {Boffetta}, \citenamefont {Marchetti},\ and\
  \citenamefont {Eckhardt}}]{Linkmannetal_PRL_2019}%
  \BibitemOpen
  \bibfield  {author} {\bibinfo {author} {\bibfnamefont {M.}~\bibnamefont
  {Linkmann}}, \bibinfo {author} {\bibfnamefont {G.}~\bibnamefont {Boffetta}},
  \bibinfo {author} {\bibfnamefont {M.~C.}\ \bibnamefont {Marchetti}},\ and\
  \bibinfo {author} {\bibfnamefont {B.}~\bibnamefont {Eckhardt}},\ }\bibfield
  {title} {\bibinfo {title} {Phase transition to large scale coherent
  structures in two-dimensional active matter turbulence},\ }\href
  {https://doi.org/10.1103/PhysRevLett.122.214503} {\bibfield  {journal}
  {\bibinfo  {journal} {Phys. Rev. Lett.}\ }\textbf {\bibinfo {volume} {122}},\
  \bibinfo {pages} {214503} (\bibinfo {year} {2019})}\BibitemShut {NoStop}%
\bibitem [{\citenamefont {Alert}\ \emph {et~al.}(2022)\citenamefont {Alert},
  \citenamefont {Casademunt},\ and\ \citenamefont
  {Joanny}}]{Alertetal_ARCMP_2022}%
  \BibitemOpen
  \bibfield  {author} {\bibinfo {author} {\bibfnamefont {R.}~\bibnamefont
  {Alert}}, \bibinfo {author} {\bibfnamefont {J.}~\bibnamefont {Casademunt}},\
  and\ \bibinfo {author} {\bibfnamefont {J.-F.}\ \bibnamefont {Joanny}},\
  }\bibfield  {title} {\bibinfo {title} {Active turbulence},\ }\href
  {https://doi.org/10.1146/annurev-conmatphys-082321-035957} {\bibfield
  {journal} {\bibinfo  {journal} {Ann. Rev. Cond. Matt. Phys.}\ }\textbf
  {\bibinfo {volume} {13}},\ \bibinfo {pages} {143} (\bibinfo {year}
  {2022})}\BibitemShut {NoStop}%
\bibitem [{\citenamefont {Bratanov}\ \emph {et~al.}(2015)\citenamefont
  {Bratanov}, \citenamefont {Jenko},\ and\ \citenamefont
  {Frey}}]{Bratanovetal_PNAS_2015}%
  \BibitemOpen
  \bibfield  {author} {\bibinfo {author} {\bibfnamefont {V.}~\bibnamefont
  {Bratanov}}, \bibinfo {author} {\bibfnamefont {F.}~\bibnamefont {Jenko}},\
  and\ \bibinfo {author} {\bibfnamefont {E.}~\bibnamefont {Frey}},\ }\bibfield
  {title} {\bibinfo {title} {New class of turbulence in active fluids},\ }\href
  {https://doi.org/10.1073/pnas.1509304112} {\bibfield  {journal} {\bibinfo
  {journal} {Proc. Nat. Acad. Sci. (USA)}\ }\textbf {\bibinfo {volume} {112}},\
  \bibinfo {pages} {15048} (\bibinfo {year} {2015})}\BibitemShut {NoStop}%
\bibitem [{\citenamefont {Mukherjee}\ \emph {et~al.}(2023)\citenamefont
  {Mukherjee}, \citenamefont {Singh}, \citenamefont {James},\ and\
  \citenamefont {Ray}}]{Mukherjeeetal_NP_2023}%
  \BibitemOpen
  \bibfield  {author} {\bibinfo {author} {\bibfnamefont {S.}~\bibnamefont
  {Mukherjee}}, \bibinfo {author} {\bibfnamefont {R.~K.}\ \bibnamefont
  {Singh}}, \bibinfo {author} {\bibfnamefont {M.}~\bibnamefont {James}},\ and\
  \bibinfo {author} {\bibfnamefont {S.~S.}\ \bibnamefont {Ray}},\ }\bibfield
  {title} {\bibinfo {title} {Intermittency, fluctuations and maximal chaos in
  an emergent universal state of active turbulence},\ }\href
  {https://doi.org/10.1038/s41567-023-01990-z} {\bibfield  {journal} {\bibinfo
  {journal} {Nature Phys.}\ }\textbf {\bibinfo {volume} {19}},\ \bibinfo
  {pages} {891} (\bibinfo {year} {2023})}\BibitemShut {NoStop}%
\bibitem [{\citenamefont {Kolmogorov}(1941)}]{Kolmogorov}%
  \BibitemOpen
  \bibfield  {author} {\bibinfo {author} {\bibfnamefont {A.~N.}\ \bibnamefont
  {Kolmogorov}},\ }\bibfield  {title} {\bibinfo {title} {The local structure of
  turbulence in incompressible viscous fluid for very large {Reynolds}
  numbers},\ }\href@noop {} {\bibfield  {journal} {\bibinfo  {journal} {Doklady
  Akademii Nauk SSSR}\ }\textbf {\bibinfo {volume} {30}},\ \bibinfo {pages}
  {301} (\bibinfo {year} {1941})}\BibitemShut {NoStop}%
\bibitem [{\citenamefont {Giomi}(2015)}]{Giomi_PRX_2015}%
  \BibitemOpen
  \bibfield  {author} {\bibinfo {author} {\bibfnamefont {L.}~\bibnamefont
  {Giomi}},\ }\bibfield  {title} {\bibinfo {title} {Geometry and topology of
  turbulence in active nematics},\ }\href
  {https://doi.org/10.1103/PhysRevX.5.031003} {\bibfield  {journal} {\bibinfo
  {journal} {Phys. Rev. X}\ }\textbf {\bibinfo {volume} {5}},\ \bibinfo {pages}
  {031003} (\bibinfo {year} {2015})}\BibitemShut {NoStop}%
\bibitem [{\citenamefont {Alert}\ \emph {et~al.}(2020)\citenamefont {Alert},
  \citenamefont {Joanny},\ and\ \citenamefont
  {Casademunt}}]{Alertetal_NP_2020}%
  \BibitemOpen
  \bibfield  {author} {\bibinfo {author} {\bibfnamefont {R.}~\bibnamefont
  {Alert}}, \bibinfo {author} {\bibfnamefont {J.-F.}\ \bibnamefont {Joanny}},\
  and\ \bibinfo {author} {\bibfnamefont {J.}~\bibnamefont {Casademunt}},\
  }\bibfield  {title} {\bibinfo {title} {Universal scaling of active nematic
  turbulence},\ }\href {https://doi.org/10.1038/s41567-020-0854-4} {\bibfield
  {journal} {\bibinfo  {journal} {Nature Phys.}\ }\textbf {\bibinfo {volume}
  {16}},\ \bibinfo {pages} {682} (\bibinfo {year} {2020})}\BibitemShut
  {NoStop}%
\bibitem [{\citenamefont {Xia}\ \emph {et~al.}(2013)\citenamefont {Xia},
  \citenamefont {Francois}, \citenamefont {Punzmann},\ and\ \citenamefont
  {Shats}}]{Xiaetal_NC_2013}%
  \BibitemOpen
  \bibfield  {author} {\bibinfo {author} {\bibfnamefont {H.}~\bibnamefont
  {Xia}}, \bibinfo {author} {\bibfnamefont {N.}~\bibnamefont {Francois}},
  \bibinfo {author} {\bibfnamefont {H.}~\bibnamefont {Punzmann}},\ and\
  \bibinfo {author} {\bibfnamefont {M.}~\bibnamefont {Shats}},\ }\bibfield
  {title} {\bibinfo {title} {Lagrangian scale of particle dispersion in
  turbulence},\ }\href {https://doi.org/10.1038/ncomms3013} {\bibfield
  {journal} {\bibinfo  {journal} {Nature Commun.}\ }\textbf {\bibinfo {volume}
  {4}},\ \bibinfo {pages} {2013} (\bibinfo {year} {2013})}\BibitemShut
  {NoStop}%
\bibitem [{\citenamefont {Kurtuldu}\ \emph {et~al.}(2011)\citenamefont
  {Kurtuldu}, \citenamefont {Guasto}, \citenamefont {Johnson},\ and\
  \citenamefont {Gollub}}]{Kurtulduetal_PNAS_2011}%
  \BibitemOpen
  \bibfield  {author} {\bibinfo {author} {\bibfnamefont {H.}~\bibnamefont
  {Kurtuldu}}, \bibinfo {author} {\bibfnamefont {J.~S.}\ \bibnamefont
  {Guasto}}, \bibinfo {author} {\bibfnamefont {K.~A.}\ \bibnamefont
  {Johnson}},\ and\ \bibinfo {author} {\bibfnamefont {J.~P.}\ \bibnamefont
  {Gollub}},\ }\bibfield  {title} {\bibinfo {title} {Enhancement of biomixing
  by swimming algal cells in two-dimensional films},\ }\href
  {https://doi.org/10.1073/pnas.1107046108} {\bibfield  {journal} {\bibinfo
  {journal} {Proc. Nat. Acad. Sci. (USA)}\ }\textbf {\bibinfo {volume} {108}},\
  \bibinfo {pages} {10391} (\bibinfo {year} {2011})}\BibitemShut {NoStop}%
\bibitem [{\citenamefont {Morozov}\ and\ \citenamefont
  {Marenduzzo}(2014)}]{Morozovetal_SM_2014}%
  \BibitemOpen
  \bibfield  {author} {\bibinfo {author} {\bibfnamefont {A.}~\bibnamefont
  {Morozov}}\ and\ \bibinfo {author} {\bibfnamefont {D.}~\bibnamefont
  {Marenduzzo}},\ }\bibfield  {title} {\bibinfo {title} {Enhanced diffusion of
  tracer particles in dilute bacterial suspensions},\ }\href
  {https://doi.org/10.1039/c3sm52201f} {\bibfield  {journal} {\bibinfo
  {journal} {Soft Matter}\ }\textbf {\bibinfo {volume} {10}},\ \bibinfo {pages}
  {2748} (\bibinfo {year} {2014})}\BibitemShut {NoStop}%
\bibitem [{\citenamefont {Ariel}\ \emph {et~al.}(2015)\citenamefont {Ariel},
  \citenamefont {Rabani}, \citenamefont {Benisty}, \citenamefont {Partridge},
  \citenamefont {Harshey},\ and\ \citenamefont {Be'er}}]{Arieletal_NC_2015}%
  \BibitemOpen
  \bibfield  {author} {\bibinfo {author} {\bibfnamefont {G.}~\bibnamefont
  {Ariel}}, \bibinfo {author} {\bibfnamefont {A.}~\bibnamefont {Rabani}},
  \bibinfo {author} {\bibfnamefont {S.}~\bibnamefont {Benisty}}, \bibinfo
  {author} {\bibfnamefont {J.~D.}\ \bibnamefont {Partridge}}, \bibinfo {author}
  {\bibfnamefont {R.~M.}\ \bibnamefont {Harshey}},\ and\ \bibinfo {author}
  {\bibfnamefont {A.}~\bibnamefont {Be'er}},\ }\bibfield  {title} {\bibinfo
  {title} {Swarming bacteria migrate by {Levy} walk},\ }\href
  {https://doi.org/10.1038/ncomms9396} {\bibfield  {journal} {\bibinfo
  {journal} {Nature Commun.}\ }\textbf {\bibinfo {volume} {6}},\ \bibinfo
  {pages} {8396} (\bibinfo {year} {2015})}\BibitemShut {NoStop}%
\bibitem [{\citenamefont {Mukherjee}\ \emph {et~al.}(2021)\citenamefont
  {Mukherjee}, \citenamefont {Singh}, \citenamefont {James},\ and\
  \citenamefont {Ray}}]{Mukherjeeetal_PRL_2021}%
  \BibitemOpen
  \bibfield  {author} {\bibinfo {author} {\bibfnamefont {S.}~\bibnamefont
  {Mukherjee}}, \bibinfo {author} {\bibfnamefont {R.~K.}\ \bibnamefont
  {Singh}}, \bibinfo {author} {\bibfnamefont {M.}~\bibnamefont {James}},\ and\
  \bibinfo {author} {\bibfnamefont {S.~S.}\ \bibnamefont {Ray}},\ }\bibfield
  {title} {\bibinfo {title} {Anomalous diffusion and levy walks distinguish
  active from inertial turbulence},\ }\href
  {https://doi.org/10.1103/PhysRevLett.127.118001} {\bibfield  {journal}
  {\bibinfo  {journal} {Phys. Rev. Lett.}\ }\textbf {\bibinfo {volume} {127}},\
  \bibinfo {pages} {118001} (\bibinfo {year} {2021})}\BibitemShut {NoStop}%
\bibitem [{\citenamefont {Singh}\ \emph {et~al.}(2022)\citenamefont {Singh},
  \citenamefont {Mukherjee},\ and\ \citenamefont {Ray}}]{Singhetal_PRF_2022}%
  \BibitemOpen
  \bibfield  {author} {\bibinfo {author} {\bibfnamefont {R.~K.}\ \bibnamefont
  {Singh}}, \bibinfo {author} {\bibfnamefont {S.}~\bibnamefont {Mukherjee}},\
  and\ \bibinfo {author} {\bibfnamefont {S.~S.}\ \bibnamefont {Ray}},\
  }\bibfield  {title} {\bibinfo {title} {Lagrangian manifestation of anomalies
  in active turbulence},\ }\href
  {https://doi.org/10.1103/PhysRevFluids.7.033101} {\bibfield  {journal}
  {\bibinfo  {journal} {Phys. Rev. Fluids}\ }\textbf {\bibinfo {volume} {7}},\
  \bibinfo {pages} {033101} (\bibinfo {year} {2022})}\BibitemShut {NoStop}%
\bibitem [{\citenamefont {Viswanathan}\ \emph {et~al.}(1999)\citenamefont
  {Viswanathan}, \citenamefont {Buldyrev}, \citenamefont {Havlin},
  \citenamefont {da~Luz}, \citenamefont {Raposo},\ and\ \citenamefont
  {Stanley}}]{Viswanathanetal_Nature_1999}%
  \BibitemOpen
  \bibfield  {author} {\bibinfo {author} {\bibfnamefont {G.~M.}\ \bibnamefont
  {Viswanathan}}, \bibinfo {author} {\bibfnamefont {S.~V.}\ \bibnamefont
  {Buldyrev}}, \bibinfo {author} {\bibfnamefont {S.}~\bibnamefont {Havlin}},
  \bibinfo {author} {\bibfnamefont {M.~G.~E.}\ \bibnamefont {da~Luz}}, \bibinfo
  {author} {\bibfnamefont {E.~P.}\ \bibnamefont {Raposo}},\ and\ \bibinfo
  {author} {\bibfnamefont {H.~E.}\ \bibnamefont {Stanley}},\ }\bibfield
  {title} {\bibinfo {title} {Optimizing the success of random searches},\
  }\href {https://doi.org/10.1038/44831} {\bibfield  {journal} {\bibinfo
  {journal} {Nature}\ }\textbf {\bibinfo {volume} {401}},\ \bibinfo {pages}
  {911} (\bibinfo {year} {1999})}\BibitemShut {NoStop}%
\bibitem [{\citenamefont {Bartumeus}\ \emph {et~al.}(2002)\citenamefont
  {Bartumeus}, \citenamefont {Catalan}, \citenamefont {Fulco}, \citenamefont
  {Lyra},\ and\ \citenamefont {Viswanathan}}]{Bartumeusetal_PRL_2002}%
  \BibitemOpen
  \bibfield  {author} {\bibinfo {author} {\bibfnamefont {F.}~\bibnamefont
  {Bartumeus}}, \bibinfo {author} {\bibfnamefont {J.}~\bibnamefont {Catalan}},
  \bibinfo {author} {\bibfnamefont {U.~L.}\ \bibnamefont {Fulco}}, \bibinfo
  {author} {\bibfnamefont {M.}~\bibnamefont {Lyra}},\ and\ \bibinfo {author}
  {\bibfnamefont {G.~M.}\ \bibnamefont {Viswanathan}},\ }\bibfield  {title}
  {\bibinfo {title} {Optimizing the encounter rate in biological interactions:
  Levy versus brownian strategies},\ }\href
  {https://doi.org/10.1103/PhysRevLett.88.097901} {\bibfield  {journal}
  {\bibinfo  {journal} {Phys. Rev. Lett.}\ }\textbf {\bibinfo {volume} {88}},\
  \bibinfo {pages} {097901} (\bibinfo {year} {2002})}\BibitemShut {NoStop}%
\bibitem [{\citenamefont {Humphries}\ \emph {et~al.}(2012)\citenamefont
  {Humphries}, \citenamefont {Weimerskirch}, \citenamefont {Queiroz},
  \citenamefont {Southall},\ and\ \citenamefont
  {Sims}}]{Humphriesetal_PNAS_2012}%
  \BibitemOpen
  \bibfield  {author} {\bibinfo {author} {\bibfnamefont {N.~E.}\ \bibnamefont
  {Humphries}}, \bibinfo {author} {\bibfnamefont {H.}~\bibnamefont
  {Weimerskirch}}, \bibinfo {author} {\bibfnamefont {N.}~\bibnamefont
  {Queiroz}}, \bibinfo {author} {\bibfnamefont {E.~J.}\ \bibnamefont
  {Southall}},\ and\ \bibinfo {author} {\bibfnamefont {D.~W.}\ \bibnamefont
  {Sims}},\ }\bibfield  {title} {\bibinfo {title} {Foraging success of
  biological {Levy} flights recorded in situ},\ }\href
  {https://doi.org/10.1073/pnas.1121201109} {\bibfield  {journal} {\bibinfo
  {journal} {Proc. Nat. Acad. Sci. (USA)}\ }\textbf {\bibinfo {volume} {109}},\
  \bibinfo {pages} {7169} (\bibinfo {year} {2012})}\BibitemShut {NoStop}%
\bibitem [{\citenamefont {Wensink}\ \emph {et~al.}(2012)\citenamefont
  {Wensink}, \citenamefont {Dunkel}, \citenamefont {Heidenreich}, \citenamefont
  {Drescher}, \citenamefont {Goldstein}, \citenamefont {L\"owen},\ and\
  \citenamefont {Yeomans}}]{Wensinketal_PNAS_2012}%
  \BibitemOpen
  \bibfield  {author} {\bibinfo {author} {\bibfnamefont {H.~H.}\ \bibnamefont
  {Wensink}}, \bibinfo {author} {\bibfnamefont {J.}~\bibnamefont {Dunkel}},
  \bibinfo {author} {\bibfnamefont {S.}~\bibnamefont {Heidenreich}}, \bibinfo
  {author} {\bibfnamefont {K.}~\bibnamefont {Drescher}}, \bibinfo {author}
  {\bibfnamefont {R.~E.}\ \bibnamefont {Goldstein}}, \bibinfo {author}
  {\bibfnamefont {H.}~\bibnamefont {L\"owen}},\ and\ \bibinfo {author}
  {\bibfnamefont {J.~M.}\ \bibnamefont {Yeomans}},\ }\bibfield  {title}
  {\bibinfo {title} {Meso-scale turbulence in living fluids},\ }\href
  {https://doi.org/10.1073/pnas.1202032109} {\bibfield  {journal} {\bibinfo
  {journal} {Proc. Nat. Acad. Sci. (USA)}\ }\textbf {\bibinfo {volume} {109}},\
  \bibinfo {pages} {14308} (\bibinfo {year} {2012})}\BibitemShut {NoStop}%
\bibitem [{\citenamefont {Dunkel}\ \emph {et~al.}(2013)\citenamefont {Dunkel},
  \citenamefont {Heidenreich}, \citenamefont {Drescher}, \citenamefont
  {Wensink}, \citenamefont {Baer},\ and\ \citenamefont
  {Goldstein}}]{Dunkeletal_PRL_2013}%
  \BibitemOpen
  \bibfield  {author} {\bibinfo {author} {\bibfnamefont {J.}~\bibnamefont
  {Dunkel}}, \bibinfo {author} {\bibfnamefont {S.}~\bibnamefont {Heidenreich}},
  \bibinfo {author} {\bibfnamefont {K.}~\bibnamefont {Drescher}}, \bibinfo
  {author} {\bibfnamefont {H.~H.}\ \bibnamefont {Wensink}}, \bibinfo {author}
  {\bibfnamefont {M.}~\bibnamefont {Baer}},\ and\ \bibinfo {author}
  {\bibfnamefont {R.~E.}\ \bibnamefont {Goldstein}},\ }\bibfield  {title}
  {\bibinfo {title} {Fluid dynamics of bacterial turbulence},\ }\href
  {https://doi.org/10.1103/PhysRevLett.110.228102} {\bibfield  {journal}
  {\bibinfo  {journal} {Phys. Rev. Lett.}\ }\textbf {\bibinfo {volume} {110}},\
  \bibinfo {pages} {228102} (\bibinfo {year} {2013})}\BibitemShut {NoStop}%
\bibitem [{\citenamefont {Slomka}\ and\ \citenamefont
  {Dunkel}(2017)}]{Slomkaetal_PRF_2017}%
  \BibitemOpen
  \bibfield  {author} {\bibinfo {author} {\bibfnamefont {J.}~\bibnamefont
  {Slomka}}\ and\ \bibinfo {author} {\bibfnamefont {J.}~\bibnamefont
  {Dunkel}},\ }\bibfield  {title} {\bibinfo {title} {Geometry-dependent
  viscosity reduction in sheared active fluids},\ }\href
  {https://doi.org/10.1103/PhysRevFluids.2.043102} {\bibfield  {journal}
  {\bibinfo  {journal} {Phys. Rev. Fluids}\ }\textbf {\bibinfo {volume} {2}},\
  \bibinfo {pages} {043102} (\bibinfo {year} {2017})}\BibitemShut {NoStop}%
\bibitem [{\citenamefont {Sokolov}\ and\ \citenamefont
  {Aranson}(2012)}]{Sokolovetal_PRL_2012}%
  \BibitemOpen
  \bibfield  {author} {\bibinfo {author} {\bibfnamefont {A.}~\bibnamefont
  {Sokolov}}\ and\ \bibinfo {author} {\bibfnamefont {I.~S.}\ \bibnamefont
  {Aranson}},\ }\bibfield  {title} {\bibinfo {title} {Physical properties of
  collective motion in suspensions of bacteria},\ }\href
  {https://doi.org/10.1103/PhysRevLett.109.248109} {\bibfield  {journal}
  {\bibinfo  {journal} {Phys. Rev. Lett.}\ }\textbf {\bibinfo {volume} {109}},\
  \bibinfo {pages} {248109} (\bibinfo {year} {2012})}\BibitemShut {NoStop}%
\bibitem [{\citenamefont {Linkmann}\ \emph
  {et~al.}(2020{\natexlab{a}})\citenamefont {Linkmann}, \citenamefont
  {Marchetti}, \citenamefont {Boffetta},\ and\ \citenamefont
  {Eckhardt}}]{Linkmannetal_PRE_2020}%
  \BibitemOpen
  \bibfield  {author} {\bibinfo {author} {\bibfnamefont {M.}~\bibnamefont
  {Linkmann}}, \bibinfo {author} {\bibfnamefont {M.~C.}\ \bibnamefont
  {Marchetti}}, \bibinfo {author} {\bibfnamefont {G.}~\bibnamefont
  {Boffetta}},\ and\ \bibinfo {author} {\bibfnamefont {B.}~\bibnamefont
  {Eckhardt}},\ }\bibfield  {title} {\bibinfo {title} {{Condensate formation
  and multiscale dynamics in two-dimensional active suspensions}},\ }\href
  {https://doi.org/10.1103/PhysRevE.101.022609} {\bibfield  {journal} {\bibinfo
   {journal} {Phys. Rev. E}\ }\textbf {\bibinfo {volume} {101}},\ \bibinfo
  {pages} {022609} (\bibinfo {year} {2020}{\natexlab{a}})}\BibitemShut
  {NoStop}%
\bibitem [{\citenamefont {Linkmann}\ \emph
  {et~al.}(2020{\natexlab{b}})\citenamefont {Linkmann}, \citenamefont
  {Hohmann},\ and\ \citenamefont {Eckhardt}}]{Linkmannetal_JFM_2020}%
  \BibitemOpen
  \bibfield  {author} {\bibinfo {author} {\bibfnamefont {M.}~\bibnamefont
  {Linkmann}}, \bibinfo {author} {\bibfnamefont {M.}~\bibnamefont {Hohmann}},\
  and\ \bibinfo {author} {\bibfnamefont {B.}~\bibnamefont {Eckhardt}},\
  }\bibfield  {title} {\bibinfo {title} {{Non-universal transitions to
  two-dimensional turbulence}},\ }\href {https://doi.org/10.1017/jfm.2020.198}
  {\bibfield  {journal} {\bibinfo  {journal} {J. Fluid Mech.}\ }\textbf
  {\bibinfo {volume} {892}},\ \bibinfo {pages} {A18} (\bibinfo {year}
  {2020}{\natexlab{b}})}\BibitemShut {NoStop}%
\bibitem [{\citenamefont {Boffetta}\ and\ \citenamefont
  {Ecke}(2012)}]{Boffettaetal_ARFM_2012}%
  \BibitemOpen
  \bibfield  {author} {\bibinfo {author} {\bibfnamefont {G.}~\bibnamefont
  {Boffetta}}\ and\ \bibinfo {author} {\bibfnamefont {R.~E.}\ \bibnamefont
  {Ecke}},\ }\bibfield  {title} {\bibinfo {title} {Two-dimensional
  turbulence},\ }\href {https://doi.org/10.1146/annurev-fluid-120710-101240}
  {\bibfield  {journal} {\bibinfo  {journal} {Ann. Rev. Fluid Mech.}\ }\textbf
  {\bibinfo {volume} {44}},\ \bibinfo {pages} {427} (\bibinfo {year}
  {2012})}\BibitemShut {NoStop}%
\bibitem [{\citenamefont {Ilkanaiv}\ \emph {et~al.}(2017)\citenamefont
  {Ilkanaiv}, \citenamefont {Kearns}, \citenamefont {Ariel},\ and\
  \citenamefont {Be'er}}]{Ilkanaivetal_PRL_2017}%
  \BibitemOpen
  \bibfield  {author} {\bibinfo {author} {\bibfnamefont {B.}~\bibnamefont
  {Ilkanaiv}}, \bibinfo {author} {\bibfnamefont {D.~B.}\ \bibnamefont
  {Kearns}}, \bibinfo {author} {\bibfnamefont {G.}~\bibnamefont {Ariel}},\ and\
  \bibinfo {author} {\bibfnamefont {A.}~\bibnamefont {Be'er}},\ }\bibfield
  {title} {\bibinfo {title} {Effect of cell aspect ratio on swarming
  bacteria},\ }\href {https://doi.org/10.1103/PhysRevLett.118.158002}
  {\bibfield  {journal} {\bibinfo  {journal} {Phys. Rev. Lett.}\ }\textbf
  {\bibinfo {volume} {118}},\ \bibinfo {pages} {158002} (\bibinfo {year}
  {2017})}\BibitemShut {NoStop}%
\bibitem [{\citenamefont {Wu}\ and\ \citenamefont
  {Libchaber}(2000)}]{Wuetal_PRL_2000}%
  \BibitemOpen
  \bibfield  {author} {\bibinfo {author} {\bibfnamefont {X.}~\bibnamefont
  {Wu}}\ and\ \bibinfo {author} {\bibfnamefont {A.}~\bibnamefont {Libchaber}},\
  }\bibfield  {title} {\bibinfo {title} {Particle diffusion in a
  quasi-two-dimensional bacterial bath},\ }\href
  {https://doi.org/10.1103/PhysRevLett.84.3017} {\bibfield  {journal} {\bibinfo
   {journal} {Phys. Rev. Lett.}\ }\textbf {\bibinfo {volume} {84}},\ \bibinfo
  {pages} {3017} (\bibinfo {year} {2000})}\BibitemShut {NoStop}%
\bibitem [{\citenamefont {Ma}\ and\ \citenamefont
  {Torquato}(2017)}]{Maetal_JAP_2017}%
  \BibitemOpen
  \bibfield  {author} {\bibinfo {author} {\bibfnamefont {Z.}~\bibnamefont
  {Ma}}\ and\ \bibinfo {author} {\bibfnamefont {S.}~\bibnamefont {Torquato}},\
  }\bibfield  {title} {\bibinfo {title} {Random scalar fields and
  hyperuniformity},\ }\href {https://doi.org/10.1063/1.4989492} {\bibfield
  {journal} {\bibinfo  {journal} {J. Appl. Phys.}\ }\textbf {\bibinfo {volume}
  {121}},\ \bibinfo {pages} {244904} (\bibinfo {year} {2017})}\BibitemShut
  {NoStop}%
\bibitem [{\citenamefont {Torquato}(2016)}]{Tor16}%
  \BibitemOpen
  \bibfield  {author} {\bibinfo {author} {\bibfnamefont {S.}~\bibnamefont
  {Torquato}},\ }\bibfield  {title} {\bibinfo {title} {{Hyperuniformity and its
  generalizations}},\ }\href {https://doi.org/10.1103/PhysRevE.94.022122}
  {\bibfield  {journal} {\bibinfo  {journal} {Physical Review E}\ }\textbf
  {\bibinfo {volume} {94}},\ \bibinfo {pages} {1} (\bibinfo {year} {2016})},\
  \Eprint {https://arxiv.org/abs/1607.08814} {arXiv:1607.08814} \BibitemShut
  {NoStop}%
\bibitem [{\citenamefont {Salvalaglio}\ \emph {et~al.}(2020)\citenamefont
  {Salvalaglio}, \citenamefont {Bouabdellaoui}, \citenamefont {Bollani},
  \citenamefont {Benali}, \citenamefont {Favre}, \citenamefont {Claude},
  \citenamefont {Wenger}, \citenamefont {de~Anna}, \citenamefont {Intonti},
  \citenamefont {Voigt} \emph {et~al.}}]{salvalaglio2020hyperuniform}%
  \BibitemOpen
  \bibfield  {author} {\bibinfo {author} {\bibfnamefont {M.}~\bibnamefont
  {Salvalaglio}}, \bibinfo {author} {\bibfnamefont {M.}~\bibnamefont
  {Bouabdellaoui}}, \bibinfo {author} {\bibfnamefont {M.}~\bibnamefont
  {Bollani}}, \bibinfo {author} {\bibfnamefont {A.}~\bibnamefont {Benali}},
  \bibinfo {author} {\bibfnamefont {L.}~\bibnamefont {Favre}}, \bibinfo
  {author} {\bibfnamefont {J.-B.}\ \bibnamefont {Claude}}, \bibinfo {author}
  {\bibfnamefont {J.}~\bibnamefont {Wenger}}, \bibinfo {author} {\bibfnamefont
  {P.}~\bibnamefont {de~Anna}}, \bibinfo {author} {\bibfnamefont
  {F.}~\bibnamefont {Intonti}}, \bibinfo {author} {\bibfnamefont
  {A.}~\bibnamefont {Voigt}}, \emph {et~al.},\ }\bibfield  {title} {\bibinfo
  {title} {Hyperuniform monocrystalline structures by spinodal solid-state
  dewetting},\ }\href@noop {} {\bibfield  {journal} {\bibinfo  {journal} {Phys.
  Rev. Lett}\ }\textbf {\bibinfo {volume} {125}},\ \bibinfo {pages} {126101}
  (\bibinfo {year} {2020})}\BibitemShut {NoStop}%
\bibitem [{\citenamefont {Torquato}(1999)}]{Torquato1999}%
  \BibitemOpen
  \bibfield  {author} {\bibinfo {author} {\bibfnamefont {S.}~\bibnamefont
  {Torquato}},\ }\bibfield  {title} {\bibinfo {title} {{Exact conditions on
  physically realizable correlation functions of random media}},\ }\href
  {https://doi.org/10.1063/1.480255} {\bibfield  {journal} {\bibinfo  {journal}
  {The Journal of Chemical Physics}\ }\textbf {\bibinfo {volume} {111}},\
  \bibinfo {pages} {8832} (\bibinfo {year} {1999})}\BibitemShut {NoStop}%
\bibitem [{\citenamefont {Peng}\ \emph {et~al.}(2021)\citenamefont {Peng},
  \citenamefont {Liu},\ and\ \citenamefont {Cheng}}]{Pengetal_SA_2021}%
  \BibitemOpen
  \bibfield  {author} {\bibinfo {author} {\bibfnamefont {Y.}~\bibnamefont
  {Peng}}, \bibinfo {author} {\bibfnamefont {Z.}~\bibnamefont {Liu}},\ and\
  \bibinfo {author} {\bibfnamefont {X.}~\bibnamefont {Cheng}},\ }\bibfield
  {title} {\bibinfo {title} {Imaging the emergence of bacterial turbulence:
  Phase diagram and transition kinetics},\ }\href
  {https://doi.org/10.1126/sciadv.abd1240} {\bibfield  {journal} {\bibinfo
  {journal} {Science Advances}\ }\textbf {\bibinfo {volume} {7}},\ \bibinfo
  {pages} {eabd1240} (\bibinfo {year} {2021})}\BibitemShut {NoStop}%
\bibitem [{\citenamefont {Torquato}\ \emph {et~al.}(2006)\citenamefont
  {Torquato}, \citenamefont {Uche},\ and\ \citenamefont
  {Stillinger}}]{Torquato2006}%
  \BibitemOpen
  \bibfield  {author} {\bibinfo {author} {\bibfnamefont {S.}~\bibnamefont
  {Torquato}}, \bibinfo {author} {\bibfnamefont {O.~U.}\ \bibnamefont {Uche}},\
  and\ \bibinfo {author} {\bibfnamefont {F.~H.}\ \bibnamefont {Stillinger}},\
  }\bibfield  {title} {\bibinfo {title} {Random sequential addition of hard
  spheres in high euclidean dimensions},\ }\href
  {https://doi.org/10.1103/PhysRevE.74.061308} {\bibfield  {journal} {\bibinfo
  {journal} {Phys. Rev. E}\ }\textbf {\bibinfo {volume} {74}},\ \bibinfo
  {pages} {061308} (\bibinfo {year} {2006})}\BibitemShut {NoStop}%
\bibitem [{\citenamefont {Kim}\ and\ \citenamefont {Torquato}(2018)}]{Kim2018}%
  \BibitemOpen
  \bibfield  {author} {\bibinfo {author} {\bibfnamefont {J.}~\bibnamefont
  {Kim}}\ and\ \bibinfo {author} {\bibfnamefont {S.}~\bibnamefont {Torquato}},\
  }\bibfield  {title} {\bibinfo {title} {Effect of imperfections on the
  hyperuniformity of many-body systems},\ }\href
  {https://doi.org/10.1103/PhysRevB.97.054105} {\bibfield  {journal} {\bibinfo
  {journal} {Phys. Rev. B}\ }\textbf {\bibinfo {volume} {97}},\ \bibinfo
  {pages} {054105} (\bibinfo {year} {2018})}\BibitemShut {NoStop}%
\bibitem [{\citenamefont {Nishiguchi}\ and\ \citenamefont
  {Sano}(2015)}]{Nishiguchietal_PRE_2015}%
  \BibitemOpen
  \bibfield  {author} {\bibinfo {author} {\bibfnamefont {D.}~\bibnamefont
  {Nishiguchi}}\ and\ \bibinfo {author} {\bibfnamefont {M.}~\bibnamefont
  {Sano}},\ }\bibfield  {title} {\bibinfo {title} {Mesoscopic turbulence and
  local order in janus particles self-propelling under an ac electric field},\
  }\href {https://doi.org/10.1103/PhysRevE.92.052309} {\bibfield  {journal}
  {\bibinfo  {journal} {Phys. Rev. E}\ }\textbf {\bibinfo {volume} {92}},\
  \bibinfo {pages} {052309} (\bibinfo {year} {2015})}\BibitemShut {NoStop}%
\bibitem [{\citenamefont {Creppy}\ \emph {et~al.}(2015)\citenamefont {Creppy},
  \citenamefont {Praud}, \citenamefont {Druart}, \citenamefont {Kohnke},\ and\
  \citenamefont {Plourabou\'e}}]{Creppyetal_PRE_2015}%
  \BibitemOpen
  \bibfield  {author} {\bibinfo {author} {\bibfnamefont {A.}~\bibnamefont
  {Creppy}}, \bibinfo {author} {\bibfnamefont {O.}~\bibnamefont {Praud}},
  \bibinfo {author} {\bibfnamefont {X.}~\bibnamefont {Druart}}, \bibinfo
  {author} {\bibfnamefont {P.~L.}\ \bibnamefont {Kohnke}},\ and\ \bibinfo
  {author} {\bibfnamefont {F.}~\bibnamefont {Plourabou\'e}},\ }\bibfield
  {title} {\bibinfo {title} {Turbulence of swarming sperm},\ }\href
  {https://doi.org/10.1103/PhysRevE.92.032722} {\bibfield  {journal} {\bibinfo
  {journal} {Phys. Rev. E}\ }\textbf {\bibinfo {volume} {92}},\ \bibinfo
  {pages} {032722} (\bibinfo {year} {2015})}\BibitemShut {NoStop}%
\bibitem [{\citenamefont {Lin}\ \emph {et~al.}(2021)\citenamefont {Lin},
  \citenamefont {Zhang}, \citenamefont {Bi}, \citenamefont {Li},\ and\
  \citenamefont {Feng}}]{Linetal_CP_2021}%
  \BibitemOpen
  \bibfield  {author} {\bibinfo {author} {\bibfnamefont {S.-Z.}\ \bibnamefont
  {Lin}}, \bibinfo {author} {\bibfnamefont {W.-Y.}\ \bibnamefont {Zhang}},
  \bibinfo {author} {\bibfnamefont {D.}~\bibnamefont {Bi}}, \bibinfo {author}
  {\bibfnamefont {B.}~\bibnamefont {Li}},\ and\ \bibinfo {author}
  {\bibfnamefont {X.-Q.}\ \bibnamefont {Feng}},\ }\bibfield  {title} {\bibinfo
  {title} {Energetics of mesoscale cell turbulence in two-dimensional
  monolayers},\ }\href {https://doi.org/10.1038/s42005-021-00530-6} {\bibfield
  {journal} {\bibinfo  {journal} {Communications Physics}\ }\textbf {\bibinfo
  {volume} {4}},\ \bibinfo {pages} {21} (\bibinfo {year} {2021})}\BibitemShut
  {NoStop}%
\bibitem [{\citenamefont {Mart\'{\i}nez-Prat}\ \emph
  {et~al.}(2021)\citenamefont {Mart\'{\i}nez-Prat}, \citenamefont {Alert},
  \citenamefont {Meng}, \citenamefont {Ign\'es-Mullol}, \citenamefont {Joanny},
  \citenamefont {Casademunt}, \citenamefont {Golestanian},\ and\ \citenamefont
  {Sagu\'es}}]{Martinez-Pratetal_PRX_2021}%
  \BibitemOpen
  \bibfield  {author} {\bibinfo {author} {\bibfnamefont {B.}~\bibnamefont
  {Mart\'{\i}nez-Prat}}, \bibinfo {author} {\bibfnamefont {R.}~\bibnamefont
  {Alert}}, \bibinfo {author} {\bibfnamefont {F.}~\bibnamefont {Meng}},
  \bibinfo {author} {\bibfnamefont {J.}~\bibnamefont {Ign\'es-Mullol}},
  \bibinfo {author} {\bibfnamefont {J.-F.}\ \bibnamefont {Joanny}}, \bibinfo
  {author} {\bibfnamefont {J.}~\bibnamefont {Casademunt}}, \bibinfo {author}
  {\bibfnamefont {R.}~\bibnamefont {Golestanian}},\ and\ \bibinfo {author}
  {\bibfnamefont {F.}~\bibnamefont {Sagu\'es}},\ }\bibfield  {title} {\bibinfo
  {title} {Scaling regimes of active turbulence with external dissipation},\
  }\href {https://doi.org/10.1103/PhysRevX.11.031065} {\bibfield  {journal}
  {\bibinfo  {journal} {Phys. Rev. X}\ }\textbf {\bibinfo {volume} {11}},\
  \bibinfo {pages} {031065} (\bibinfo {year} {2021})}\BibitemShut {NoStop}%
\bibitem [{\citenamefont {Korobkova}\ \emph {et~al.}(2004)\citenamefont
  {Korobkova}, \citenamefont {Emonet}, \citenamefont {Vilar}, \citenamefont
  {Shimizu},\ and\ \citenamefont {Cluzel}}]{Korobkovaetal_Nature_2004}%
  \BibitemOpen
  \bibfield  {author} {\bibinfo {author} {\bibfnamefont {E.}~\bibnamefont
  {Korobkova}}, \bibinfo {author} {\bibfnamefont {T.}~\bibnamefont {Emonet}},
  \bibinfo {author} {\bibfnamefont {J.~M.~G.}\ \bibnamefont {Vilar}}, \bibinfo
  {author} {\bibfnamefont {T.~S.}\ \bibnamefont {Shimizu}},\ and\ \bibinfo
  {author} {\bibfnamefont {P.}~\bibnamefont {Cluzel}},\ }\bibfield  {title}
  {\bibinfo {title} {From molecular noise to behavioural variability in a
  single bacterium},\ }\href@noop {} {\bibfield  {journal} {\bibinfo  {journal}
  {Nature}\ }\textbf {\bibinfo {volume} {428}},\ \bibinfo {pages} {574}
  (\bibinfo {year} {2004})}\BibitemShut {NoStop}%
\bibitem [{\citenamefont {Harris}\ \emph {et~al.}(2004)\citenamefont {Harris},
  \citenamefont {Banigan}, \citenamefont {Christian}, \citenamefont {Konradt},
  \citenamefont {Wojno}, \citenamefont {Norose}, \citenamefont {Wilson},
  \citenamefont {John}, \citenamefont {Weninger}, \citenamefont {Luster},
  \citenamefont {Liu},\ and\ \citenamefont {Hunter}}]{Harrisetal_Nature_2012}%
  \BibitemOpen
  \bibfield  {author} {\bibinfo {author} {\bibfnamefont {T.~H.}\ \bibnamefont
  {Harris}}, \bibinfo {author} {\bibfnamefont {E.~J.}\ \bibnamefont {Banigan}},
  \bibinfo {author} {\bibfnamefont {D.~A.}\ \bibnamefont {Christian}}, \bibinfo
  {author} {\bibfnamefont {C.}~\bibnamefont {Konradt}}, \bibinfo {author}
  {\bibfnamefont {E.~D.~T.}\ \bibnamefont {Wojno}}, \bibinfo {author}
  {\bibfnamefont {K.}~\bibnamefont {Norose}}, \bibinfo {author} {\bibfnamefont
  {E.~H.}\ \bibnamefont {Wilson}}, \bibinfo {author} {\bibfnamefont
  {B.}~\bibnamefont {John}}, \bibinfo {author} {\bibfnamefont {W.}~\bibnamefont
  {Weninger}}, \bibinfo {author} {\bibfnamefont {A.~D.}\ \bibnamefont
  {Luster}}, \bibinfo {author} {\bibfnamefont {A.~J.}\ \bibnamefont {Liu}},\
  and\ \bibinfo {author} {\bibfnamefont {C.~A.}\ \bibnamefont {Hunter}},\
  }\bibfield  {title} {\bibinfo {title} {Generalized {Levy} walks and the role
  of chemokines in migration of effector (cd8$^+$ t) cells},\ }\href
  {https://doi.org/10.1038/nature11098} {\bibfield  {journal} {\bibinfo
  {journal} {Nature}\ }\textbf {\bibinfo {volume} {486}},\ \bibinfo {pages}
  {545} (\bibinfo {year} {2004})}\BibitemShut {NoStop}%
\bibitem [{\citenamefont {Estrada-Rodriguez}\ and\ \citenamefont
  {Perthame}(2022)}]{Estrada-Rodriguezetal_JNS_2022}%
  \BibitemOpen
  \bibfield  {author} {\bibinfo {author} {\bibfnamefont {G.}~\bibnamefont
  {Estrada-Rodriguez}}\ and\ \bibinfo {author} {\bibfnamefont {B.}~\bibnamefont
  {Perthame}},\ }\bibfield  {title} {\bibinfo {title} {Motility switching and
  front-back synchronisation in polarised cells},\ }\href
  {https://doi.org/10.1007/s00332-022-09791-z} {\bibfield  {journal} {\bibinfo
  {journal} {J. Nonl. Sci.}\ }\textbf {\bibinfo {volume} {32}},\ \bibinfo
  {pages} {40} (\bibinfo {year} {2022})}\BibitemShut {NoStop}%
\bibitem [{\citenamefont {Duncan}\ \emph {et~al.}(2022)\citenamefont {Duncan},
  \citenamefont {Estrada-Rodriguez}, \citenamefont {Stocek}, \citenamefont
  {Dragone}, \citenamefont {Vargas},\ and\ \citenamefont
  {Gimperlein}}]{Duncanetal_BB_2022}%
  \BibitemOpen
  \bibfield  {author} {\bibinfo {author} {\bibfnamefont {S.}~\bibnamefont
  {Duncan}}, \bibinfo {author} {\bibfnamefont {G.}~\bibnamefont
  {Estrada-Rodriguez}}, \bibinfo {author} {\bibfnamefont {J.}~\bibnamefont
  {Stocek}}, \bibinfo {author} {\bibfnamefont {M.}~\bibnamefont {Dragone}},
  \bibinfo {author} {\bibfnamefont {P.~A.}\ \bibnamefont {Vargas}},\ and\
  \bibinfo {author} {\bibfnamefont {H.}~\bibnamefont {Gimperlein}},\ }\bibfield
   {title} {\bibinfo {title} {Efficient quantitative assessment of robot
  swarms: coverage and targeting lévy strategies},\ }\href
  {https://doi.org/10.1088/1748-3190/ac57f0} {\bibfield  {journal} {\bibinfo
  {journal} {Bioinspiration \& Biomimetics}\ }\textbf {\bibinfo {volume}
  {17}},\ \bibinfo {pages} {036006} (\bibinfo {year} {2022})}\BibitemShut
  {NoStop}%
\bibitem [{\citenamefont {Li}\ and\ \citenamefont {Meneveau}(2005)}]{LM05}%
  \BibitemOpen
  \bibfield  {author} {\bibinfo {author} {\bibfnamefont {Y.}~\bibnamefont
  {Li}}\ and\ \bibinfo {author} {\bibfnamefont {C.}~\bibnamefont {Meneveau}},\
  }\bibfield  {title} {\bibinfo {title} {{Origin of non-gaussian statistics in
  hydrodynamic turbulence}},\ }\href
  {https://doi.org/10.1103/PhysRevLett.95.164502} {\bibfield  {journal}
  {\bibinfo  {journal} {Physical Review Letters}\ }\textbf {\bibinfo {volume}
  {95}},\ \bibinfo {pages} {1} (\bibinfo {year} {2005})}\BibitemShut {NoStop}%
\bibitem [{\citenamefont {Li}\ \emph {et~al.}(2008)\citenamefont {Li},
  \citenamefont {Perlman}, \citenamefont {Wan}, \citenamefont {Yang},
  \citenamefont {Meneveau}, \citenamefont {Burns}, \citenamefont {Chen},
  \citenamefont {Szalay},\ and\ \citenamefont {Eyink}}]{LPE08}%
  \BibitemOpen
  \bibfield  {author} {\bibinfo {author} {\bibfnamefont {Y.}~\bibnamefont
  {Li}}, \bibinfo {author} {\bibfnamefont {E.}~\bibnamefont {Perlman}},
  \bibinfo {author} {\bibfnamefont {M.}~\bibnamefont {Wan}}, \bibinfo {author}
  {\bibfnamefont {Y.}~\bibnamefont {Yang}}, \bibinfo {author} {\bibfnamefont
  {C.}~\bibnamefont {Meneveau}}, \bibinfo {author} {\bibfnamefont
  {R.}~\bibnamefont {Burns}}, \bibinfo {author} {\bibfnamefont
  {S.}~\bibnamefont {Chen}}, \bibinfo {author} {\bibfnamefont {A.}~\bibnamefont
  {Szalay}},\ and\ \bibinfo {author} {\bibfnamefont {G.}~\bibnamefont
  {Eyink}},\ }\bibfield  {title} {\bibinfo {title} {{A public turbulence
  database cluster and applications to study Lagrangian evolution of velocity
  increments in turbulence}},\ }\href
  {https://doi.org/10.1080/14685240802376389} {\bibfield  {journal} {\bibinfo
  {journal} {Journal of Turbulence}\ }\textbf {\bibinfo {volume} {9}},\
  \bibinfo {pages} {1} (\bibinfo {year} {2008})},\ \Eprint
  {https://arxiv.org/abs/0804.1703} {arXiv:0804.1703} \BibitemShut {NoStop}%
\bibitem [{\citenamefont {Yakhot}\ and\ \citenamefont {Donzis}(2018)}]{YD18}%
  \BibitemOpen
  \bibfield  {author} {\bibinfo {author} {\bibfnamefont {V.}~\bibnamefont
  {Yakhot}}\ and\ \bibinfo {author} {\bibfnamefont {D.~A.}\ \bibnamefont
  {Donzis}},\ }\bibfield  {title} {\bibinfo {title} {{Anomalous exponents in
  strong turbulence}},\ }\href {https://doi.org/10.1016/j.physd.2018.07.005}
  {\bibfield  {journal} {\bibinfo  {journal} {Physica D: Nonlinear Phenomena}\
  }\textbf {\bibinfo {volume} {384-385}},\ \bibinfo {pages} {12} (\bibinfo
  {year} {2018})}\BibitemShut {NoStop}%
\end{thebibliography}%

\end{document}